\documentclass[aps,twocolumn,pra,superscriptaddress,amsmath,longbibliography]{revtex4}
\usepackage{slashed,bbm,hyperref,amsmath,amssymb,graphicx,array}
\usepackage{float}
\usepackage{breqn}
\graphicspath{{figure/}}
\usepackage{subfig,xcolor,epstopdf,inputenc}
\usepackage{placeins,afterpage} 
\allowdisplaybreaks[1-4]
\usepackage{lipsum}
\newcommand{\sig}{\sigma}
\newcommand{\calh}{\mathcal{H}}
\newcommand{\beq}{\begin{equation}}
\newcommand{\eeq}{\end{equation}}
\newcommand{\bea}{\begin{eqnarray}}
\newcommand{\eea}{\end{eqnarray}}
\newcommand{\til}{\widetilde}

\newcommand{\dg}{\dagger}

\newcommand{\bsplit}{\begin{split}}

\makeatletter
\let\cat@comma@active\@empty
\makeatother
\begin{document}
\title{Excitation Energy Transfer under Strong Laser Drive}
\author{Xuanhua Wang}
\affiliation{Department of Physics and Astronomy, Stony Brook University, Stony Brook, NY 11794, USA}
\author{Zhedong Zhang}
\email[]{To whom correspondence should be addressed. Email: zzhan26@cityu.edu.hk  or  jin.wang1@stonybrook.edu}
\affiliation{Department of Physics, City University of Hong Kong, Kowloon, Hong Kong SAR}
\author{Jin Wang}
\email[]{To whom correspondence should be addressed. Email: zzhan26@cityu.edu.hk  or  jin.wang1@stonybrook.edu}
\affiliation{Department of Physics and Astronomy, Stony Brook University, Stony Brook, NY 11794, USA}
\affiliation{Department of Chemistry, Stony Brook University, Stony Brook, NY 11794, USA}
\begin{abstract}
Strong molecule-light interaction enables the control of molecular structures and dynamical processes. A model with strong laser drive is proposed to greatly enhance the intermolecular distance of resonant energy transfer, where the molecules are strongly driven by an optical cavity. The optimal Rabi frequency and quantum yield of energy transfer are observed, resulting from the trade off between dipole-dipole interaction and molecule-cavity coupling. When the strong drive at certain Rabi frequency is applied, a larger spatial range of effective energy transfer and a slower decay rate with the distance compared to the F\"orster mechanism of resonant energy transfer are observed in our model. Our work sheds light on spectroscopic study of the cooperative energy transfer in molecular polaritons.
\end{abstract}
\maketitle
\section{Introduction}
Long-range excitation energy transfer in molecules plays a key role in the study of molecular aggregates and proteins, which involve the exciton migration over 10-100 nm. This offers a highly sensitive approach to investigate intra- and inter-molecular distances on the nanometer scale, revealing the dynamical information about biomolecular structures and interactions \cite{Hellenkamp_NatMethod2018,Lerner_Science2018,Roy_NatMethod2008}. The long-range energy transfer, however, is difficult because the efficiency dramatically drops as the donor-acceptor separation grows, resulting in the challenge for the fluorescence detection above 10 nm \cite{Schuler_COSB2013,Hevekerl_JPCB2011}. Some prominent mechanisms are responsible for this bottleneck: disorder, local trapping, vibrations and the technical difficulties in the molecule synthesis. Elaborate experiments demonstrate the megamolecular structure incorporating fluorescent proteins as a promising building block for studying the distance-dependent energy transfer in large molecules \cite{Taylor_JACS2018}. New donor-acceptor constructs and several strategies to extend the energy transfer distance have been investigated, using lanthanides \cite{Selvin_PNAS1994,Hildebrandt_CCR2014,Selvin_ARBBS2002}, quantum dots \cite{Hildebrandt_CR2017,Guo_ACSNano2019}, gold nanoparticle quenchers \cite{Yun_JACS2005,Riskowski_ACSNano2016,Samanta_NL2014} and metal-induced energy transfer \cite{Chizhik_NP2014,Ghosh_NP2019}. Although these advanced techniques have the drawbacks of further complicating the sample structures and the preparation, the gold nanoparticles still have the advantage in many applications due to the local field enhancement.

From the quantum mechanical treatment, it is well known that the quantum yield of resonant energy transfer (RET) is strongly determined by the overlap between density of the states in the donor and the acceptor, apart from the dependence on the molecular structures \cite{Mukamel_CR2009,Scholes_ARPC2003,Wang_PRA2019}. This has been extensively explored in the F\"orster theory of resonance energy transfer (FRET), treating the dipole-dipole interactions as perturbations \cite{Forster_AP1948}. Nevertheless, it is worth noting Purcell’s effect such that the photonic environment can modify the fluorescence emission of molecules, due to the drastic change of the photonic spectral density in confined geometry \cite{Purcell_PR1946,Agarwal_book2012}. This leads to the so-called molecular polaritons that opens a new avenue to control the resonant energy transfer. Recently the functions of molecules in strong coupling to the light draw much attention, and the resonant energy transfer is one of the most extensively studied topics \cite{Zhang_JPCL2019, Dorfman_PNAS2018, Zhou_CS2018, Vidal_PRB2018}. Significant efforts were devoted to the microcavties \cite{Andrew_Science2000, Rustomji_PRX2019, Schleifenbaum_PCCP2014}, surface plasmon polaritons \cite{Hsu_JPCL2017, Andrew_Science2004}, nano-apertures \cite{Fore_NL2007, Torres_CPC2015, Chen_PNAS2014, Goldschen_AC2017, Baibakov_ACSO2020}, nanoparticles \cite{Zhang_JPCC2007, Reil_NL2008, Lunz_NL2011, Pustovit_PRB2011, Zhao_JPCC2012, Gonzaga_JCP2013,Bohlen_NS2019} and hyperbolic metameterials \cite{Tumkur_FD2015, Roth_ACSP2018}, in attempts to overcome the 10nm barrier in diffraction-limited confocal microscopes so that to enhance the quantum yield of FRET. Polaritons can act as an efficient and ultrafast energy-transfer pathway between exciton states, evident by the photoluminescence spectroscopy of J-aggregated molecular dyes placed in an optical cavity \cite{Coles_NatMat2014}. However, FRET assumes the weak dipole-dipole interaction to make the perturbation applicable, which is no longer exact in molecular polaritons since the joint photon-matter states may result in a large density of states overlap between donor and acceptor. Such drawback also exists in the transition state theory of electron transfer, despite that it does offer a microscopic description for a broad range of reaction processes \cite{Markus_JCP1956}. On the other hand, considerable enhancement of RET requires a delicate balance between exciton migration and the other donor radiative as well as nonradiative processes \cite{Bohlen_NS2019,Roth_ACSP2018,Bidault_ACSP2016}. Along this line, the remote energy transfer between molecules beyond Forster mechanism is of fundamental importance and great interest. These are still open issues, which requires significant theoretical efforts beyond the perturbative framework.

\begin{figure*}[ht]
\centering
\includegraphics[width=1 \columnwidth]{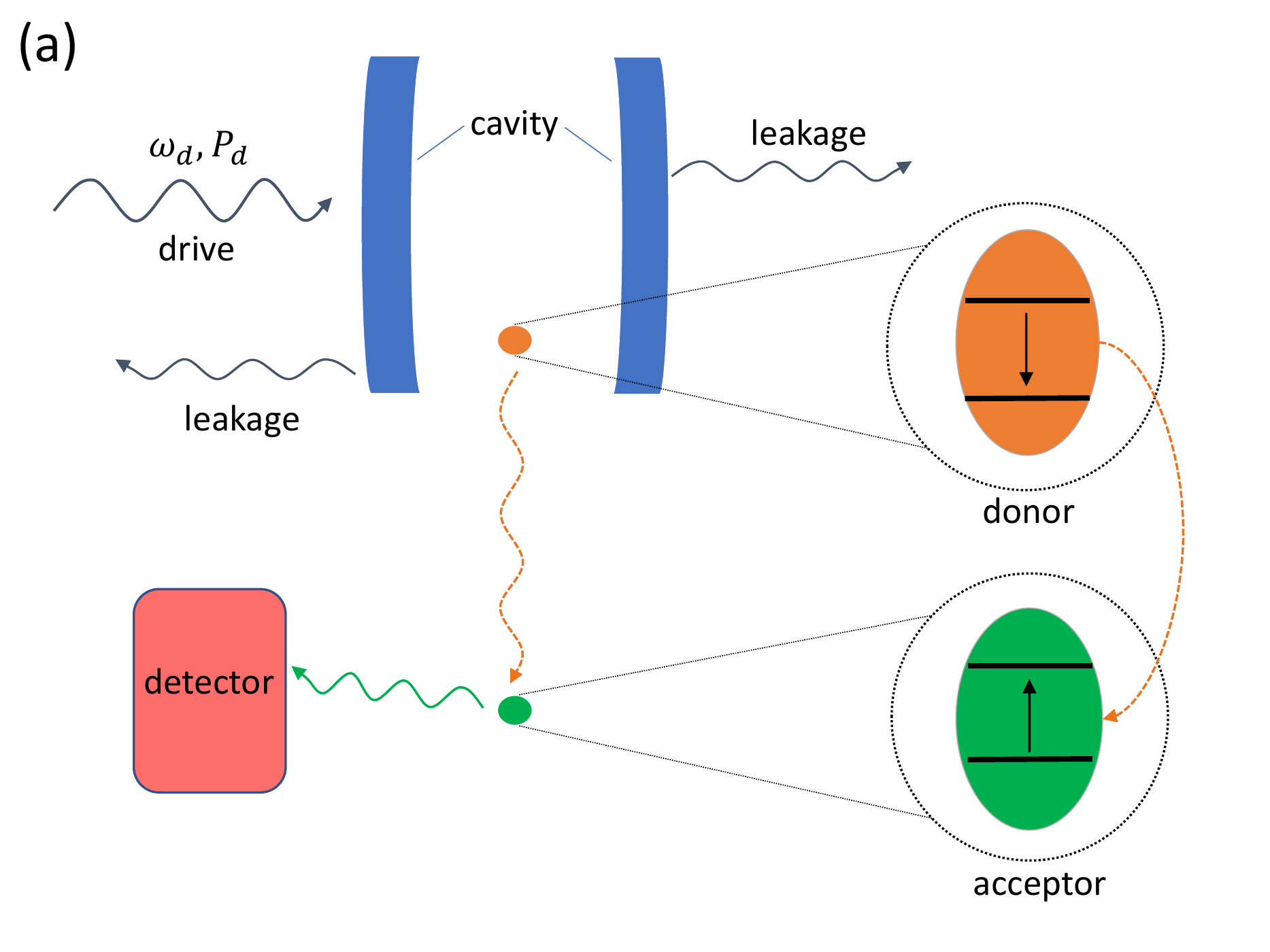}
\quad
\includegraphics[width=1 \columnwidth]{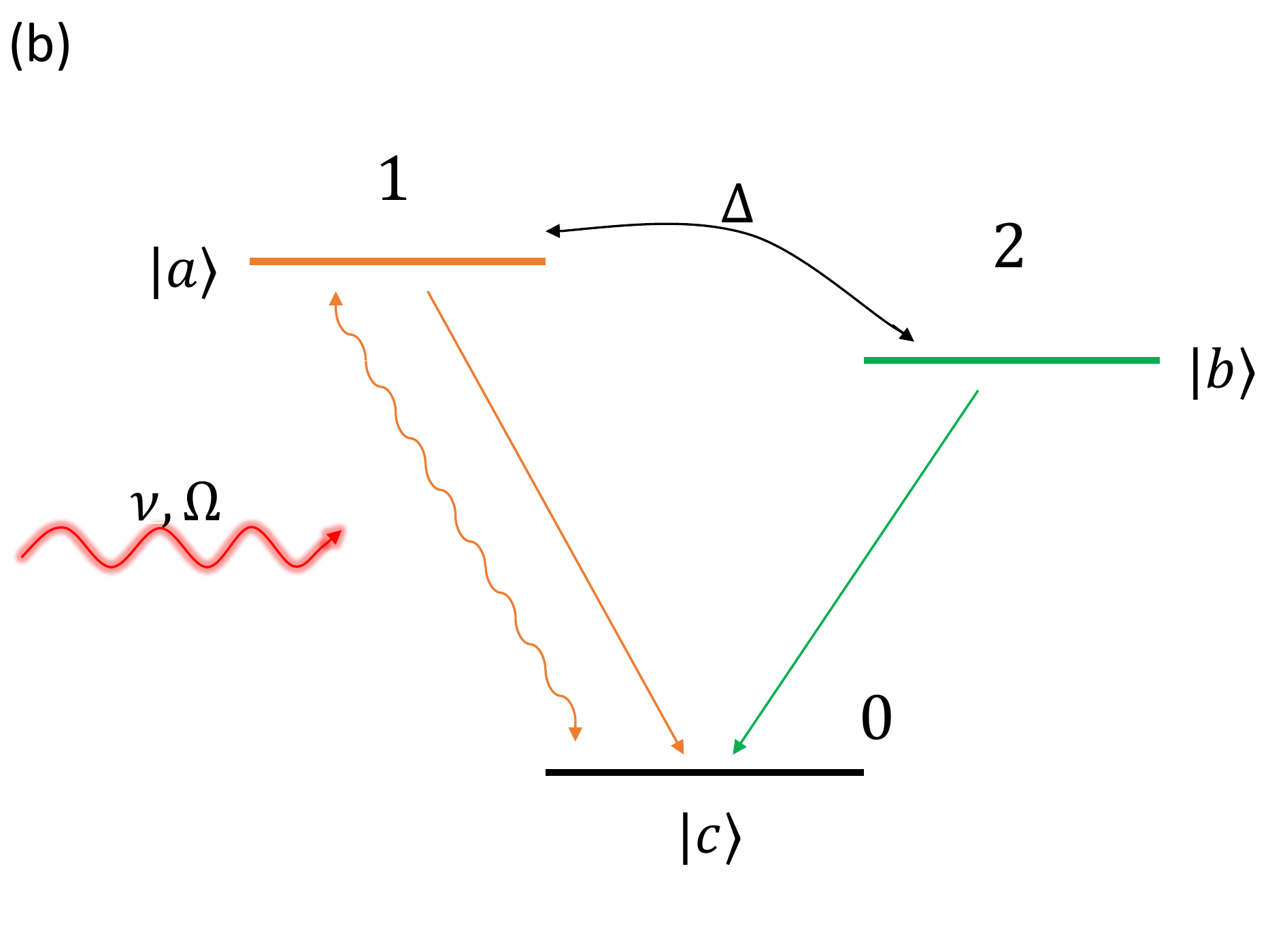}

\caption{(a) Schematic demonstration of the donor-acceptor system. (b) Schematic illustration of the simplified model of a three-level system.}
    \label{theme}
\end{figure*}

In this article we propose a long-range resonant energy transfer in molecules driven by an optical cavity. We consider such a model that the donor molecule is placed in the cavity whereas the acceptor molecule is outside. The quantum master equation incorporating the polariton effect is obtained, going beyond the perturbative treatment of dipole-dipole interaction. We observe an optimal intermolecular distance of energy transfer, showing a great enhancement through the balance between dipole-dipole interaction and the Rabi splitting induced by the joint molecule-cavity states. Our model is promising of being extended to many molecular ensembles incorporating the cooperativity, which is essential for the nonlinear spectroscopic study of cooperative energy transfer in molecular polaritons.


\section{Molecular model and equation of motion}
The schematic of energy transfer between coupled donors and acceptors is shown in Fig.~\ref{theme} (a). The donor (the orange dot) is placed inside the single-mode cavity of frequency $\nu$, which is driven by a strong laser of frequency $\omega_d$. The excitation of the donor can be transmitted through a non-radiative process to the acceptor (the green dot) outside the cavity. The cavity is implemented to amplify the photon-molecule interaction and confine the laser beam to the vicinity of the donor molecule as the wavelength of the beam is significantly longer than the distance between the donor and the acceptor molecules. The axis of the cavity is perpendicular to the displacement vector joining the donor and the acceptor, and therefore, the influence of the cavity on the energy transfer between the two molecules can be neglected. The excitation of the acceptor can be measured through a detector through its spontaneous decay. Here we assume that the field inside cavity is single mode and can be described classically when the driving laser is intense. The donor is then pumped effectively by a classical field, whose intensity varies with the driving laser. We simplify the model through ignoring the doubly-excited state due to the mismatch of transition energies caused by exciton-exciton scattering, and the schematic of the simplified system is shown in Fig.~\ref{theme} (b). 

In molecules, the vibrational motions may lead to the excited-state relaxation and dephasing. Including the vibrational degrees of freedom, under Born-Oppenheimer approximation, the full Hamiltonian of the molecular system in Fig.~\ref{theme} (a) is given as follows,
\beq \begin{split}
\calh =&\ \widetilde\omega_1 |1\rangle \langle 1|+\widetilde\omega_2 |2 \rangle \langle 2| + \Delta(|1 \rangle \langle 2|+ |2 \rangle \langle 1|)+ \nu a_{c}^\dg a_{c}\\
&+\sum_{j=1}^{2}\sum_{s}{v_s^{(j)}b_s^{\left(j\right),\dag}b_s^{\left(j\right)}}+g a_{c} |1 \rangle \langle 0 | + g^* a_{c}^\dg |0 \rangle \langle 1| \\
&+i\mathcal{E} (a_c^\dg e^{-i\omega_d t}-a_c e^{i\omega_d t})\,,
\label{H_0}
\end{split}\eeq
where
\beq
\widetilde{\omega}_j =\ \omega_j-\sum_{s}\lambda_s^{(j)}v_s^{(j)}(b_s^{\left(j\right)}+b_s^{\left(j\right),\dag})\,,
\eeq
$a_{c}$ and $a_{c}^\dg$ are cavity operators, $\nu$ and $\omega_d$ represent the cavity and the laser frequencies, $b_s^{\left(j\right)}$ and $v_s^{(j)}$ denote respectively the annihilation operator and frequency for the vibrational modes associated with molecule $j$. $\lambda_s^{(j)}$ quantifies the coupling between the electronic excitations and the vibrational modes. For many condensed-phase molecules, the vibrations have a dense distribution and can be properly described by a continuous distribution of overdamped Brownian oscillators that cause the inhomogeneous dephasing and relaxation \cite{Mukamel_PRA1990}. This is incorporated in the equation of motion for the molecular density matrix shown later, through tracing out the vibrational modes as the bath.

When both molecules are in the ground states ($|0 \rangle$), the donor can absorb a photon from the cavity and excites to the state $|1\rangle$. The excitation can be transmitted to the acceptor, which turns to the state $|2 \rangle$. $\omega_i$ is the excitation energy of each molecule, and the hopping rate between the two states is denoted as $\Delta$. The amplitude of the driving field is given by $\mathcal{E}=\sqrt{P\kappa/\omega_d}$, where $P$ is the input power and $\kappa$ is the cavity decay rate. The coupling between the cavity and the donor is $g=\mathcal{P}_{ac}\sqrt{\nu/2V}$, where $\mathcal{P}_{ac}$ is the dipole moment.

The Langevin equation \cite{Aspelmeyer_RMP2014} (ignoring the noise and the back-reaction from the donor molecule) for the rotated cavity operator $\tilde a_c=a_ce^{i\omega_d t}$ reads,
\beq
\dot{\widetilde{a}}_c=-\frac{\kappa}{2}\tilde a_c+i(\omega_d-\nu)\tilde a_c+\mathcal{E}\,.
\eeq
The steady state solution for the Langevin equation is given as:
\beq \langle \tilde a_c \rangle=\frac{\mathcal{E}}{\kappa/2-i(\omega_d-\nu)}\,.\eeq 
For strong driving fields, the fluctuations can be properly ignored. We assume that $\omega_d \approx\nu$ and replace the cavity operator $a_c$ in Eqn.~\ref{H_0} by $\langle a_c\rangle=\frac{2\mathcal{E}}{\kappa}e^{-i\nu t}$. We separate out the vibrational degrees of freedom as the environment. Keeping the relevant terms, we obtain the effective system Hamiltonian,
\beq \bsplit
\calh_S= &\omega_1 |1\rangle \langle 1|+\omega_2 |2 \rangle \langle 2| + \Delta(|1 \rangle \langle 2|+ |2 \rangle \langle 1|)\\
&+\Omega |1 \rangle \langle 0 | e^{-i\nu t}+\Omega |0 \rangle \langle 1 | e^{i\nu t}\,,
 \end{split}\label{h_s}\eeq
where $\Omega$ represents the Rabi frequency of the system. The hopping rate $\Delta$ depends on the distance between the two molecules and decays as the distance enlarges. The explicit dependence is given in Sec.~\ref{section3}. 

The effective Hamiltonian for the ambient environments and the interactions are give as follows,
\beq \calh_R=\sum_{k,\sig} \omega_k  a_{k,\sig}^\dagger a_{k,\sig}+\sum_{j=1}^{2}\sum_{s}{v_s^{(j)}b_s^{\left(j\right),\dag}b_s^{\left(j\right)}}, \eeq
and 
\beq  \bsplit
\calh_{I}=&\ -\sum_{j=1}^2\sum_{s}\lambda_s^{(j)}v_s^{(1)}(b_s^{\left(j\right)}+b_s^{\left(j\right),\dag})|j\rangle \langle j|\\
& +\sum_{j=1}^2\sum_{k,\sigma} g^{j}_{k,\sig}|j\rangle \langle 0|a_{k,\sig}+ g^{*j}_{k,\sig}|0 \rangle \langle j|a^\dg_{k,\sig}\,,
\end{split} \eeq
where $\calh_R$ is the Hamiltonian for the free ambient reservoir and the vibrational modes and $\calh_I$ describes the interaction between the system and the bath and the coupling between the electronic excitations and the vibrational modes. $a_{k,\sig}^\dag$ and $a_{k,\sig}$ are the environment photon operators. The index ``$\sigma$'' in the environment operators denotes the polarization of the photons, and $g_{k,\sig}^j$ in $\calh_I$ is the coupling which is a function of dipole moment and position. 

We treat the vibrational modes in Eqn.~\ref{h_s} as the environment to be traced out. Under Born-Markov approximation the total density matrix can be decomposed into system and bath components, $\til\rho(t)=\til \rho_s(t)\otimes \til \rho_B$. Furthermore, we apply the Weisskopf-Wigner approximation, and ignore the Cauchy principal value of the integral $\int_0^\infty d\tau e^{ix\tau}=\pi\delta(x)+i\mathcal{P}_c1/x$, then the reduced equation of motion of the system is as follows,

\begin{widetext}
\beq \bsplit
\dot{\til\rho}_s(t)= i \left[ \til \rho_s(t),\til \calh_0\right]+\Big( \sum_{j,j'}\frac{\Gamma_{jj'}}{2}\big(|0\rangle \langle j|\til\rho|j'\rangle \langle 0|-\til \rho |j \rangle \langle j'|\big)+\sum_{j}{\frac{\Gamma_{inh,jj}}{2}\left(\left|\left.j\right\rangle\left\langle j\right.\right|{\widetilde{\rho}}_s\left|\left.j\right\rangle\left\langle j\right.\right|-{\widetilde{\rho}}_s\left|\left.j\right\rangle\left\langle j\right.\right|\right)+h.c.\Big)}\,,
\label{master}
\end{split} \eeq
\end{widetext}
where we have assumed no thermal excitation from the ambient baths (vacuum surroundings). The full equations of motion is given in the Appendix. $\Gamma_{inh,jj}$ quantifies the pure dephasing from vibrations that gives the inhomogeneous line broadening. As we have adopted a continuous distribution of overdamped Brownian oscillators for vibrational modes in current model, the spectral density $S_j(\omega)$ of the oscillators essentially includes a cutoff frequency $\Lambda_j$. We then find
\beq \Gamma_{inh,jj}=S_j(\Lambda_j) \coth \left(\frac{\Lambda_j}{2T}\right) \eeq
and for the Lorentzian spectrum $S\left(\omega\right)=2\lambda_j\omega\Lambda_j/(\omega^2+\Lambda_j^2)$, $\Gamma_{inh,j}=2\lambda_j T/ \Lambda_j$ in the high temperature limit ($\Lambda_j \ll  T \ll \omega_j$ such that the temperature still cannot induce the excitation in optical range), where $\lambda_j$ is the typical coupling constant between exciton and vibration.

In Eqn.~\ref{master}, the radiative decay is given by the second term on the right hand side, with the dissipation constant $\Gamma_{jj'}$ is the dissipation constant defined as follows,
\beq \Gamma_{jj'}(\nu) \equiv \frac{V}{(2\pi)^2} \sum_{\sigma}\int d^3\vec{k}\ g^{j}_{\omega_k,\sigma} g^{*j'}_{\omega_k,\sigma}\ \delta(\nu-\omega_k)\,, \label{decay} \eeq
and $\til\calh_0$ is the rotated system Hamiltonian as given below,
\beq 
\til\calh_0=\sum_{j=1}^2 \delta_j |j\rangle\langle j|+\big(\Delta |1\rangle \langle 2|+\Omega|1 \rangle \langle 0|+h.c. \big)\,.
\eeq
The details of the calculation are provided in the Appendix. The hopping rate between the two molecules and the dissipation constants are in general distance-dependent and discussed in the next section.

\section{Resonant dipole-dipole interaction and distance dependence}\label{section3}
Since the molecular size is much smaller than the wavelength of the light, we may invoke the dipole approximation that leads to \beq g^{j}_{\omega_k,\sig}=\sqrt{\omega_k/2V}\vec{\mathcal{P}}_j \cdot \hat{\epsilon}_{\mathbf{k}}^{\sig}\ \mathrm{exp}[i\mathbf{k}\cdot\mathbf{r}_j]\,,\eeq where $\mathbf{r}_i$ is the displacement vector of the $i^{th}$ dipole, $\hat{\epsilon}_{\mathbf{k}}$ is the polarization vector of the field,  and $\mathcal{P}_i$ is the dipole moment of the $i^{\mathrm{th}}$ atom \cite{Agarwal_book2012}. The individual decay constant of the $i^{th}$ dipole following definition~\ref{decay} is 
\beq \Gamma_{ii}(\nu)= \mathcal{P}_i^2\nu^3/(3\pi)\,. \eeq 
The cross dissipation constant (or collective dissipation constant) $\Gamma_{jj'}$ for $j\ne j'$ is given as follows, 
\beq 
\Gamma_{jj'} = \frac{\nu^3}{2(2\pi)^2}\int d\vec{\Omega}_{\hat{\mathbf{k}}}\sum_{\sigma} (\vec{\mathcal{P}}_j \cdot \hat{\epsilon}_{\mathbf{k}}^\sig) (\vec{\mathcal{P}}_{j'} \cdot \hat{\epsilon}_{\mathbf{k}}^{\sig*}) e^{i\nu\hat{\mathbf k}\cdot(\mathbf{r}_j-\mathbf{r}_{j'})}\,, \label{cross}
\eeq
where $d\vec{\Omega}_{\hat{\mathbf{k}}}$ is the measure of the solid angle. Notice that by Cauchy–Schwarz inequality the cross dissipation constant $\Gamma_{jj'} \le (\Gamma_{11}\Gamma_{22})^{\frac{1}{2}}$. The collective dissipation constant $\Gamma_{12}$ decays with the increasing angle differences between the two dipoles and also with the distance between them when $|\mathbf{r}_2-\mathbf{r}_1| \lesssim 1/\nu$. Following the Eqn.~\ref{cross}, the collective dissipation constant is related to the individual dissipation constants by
\beq 
\Gamma_{jj'} =f(\xi) \sqrt{\Gamma_{11}\Gamma_{22}},
\eeq 
where $f(x)$ is the real function such that $|f(x)|\le 1$, and $\xi=\nu  |\mathbf{r}_2-\mathbf{r}_1|$ is the distance between two dipoles multiplied by the laser frequency. We will refer to the dimensionless parameter $\xi\equiv\nu  |\mathbf{r}_2-\mathbf{r}_1|=d/\lambda_d$ as the distance in this paper, which measures the separation $d$ between the donor and the acceptor in terms of the laser wavelength $\lambda_d$.
Applying $\sum_\sig \epsilon^{\sig}_{\mathbf{k}}\epsilon^{\sig}_{\mathbf{k}}+\frac{\mathbf k \mathbf k}{k^2}=1$ in Eqn.~\ref{cross} where $\mathbf k\mathbf k$ is the dyadic product, the function $f(\xi)$ of distance $\xi$ for two parallel dipoles can be calculated as follows,
\beq \bsplit f(\xi)=&\frac{3}{2}\Big[\Big(\hat{\mathcal{P}}_j\cdot\hat{\mathcal{P}}_{j'}-(\hat{\mathcal{P}}_j\cdot\hat{R})(\hat{\mathcal{P}}_{j'}\cdot\hat{R})\Big) \frac{\sin\xi}{\xi}\\
&+\Big(3(\hat{\mathcal{P}}_j\cdot\hat{R})(\hat{\mathcal{P}}_{j'}\cdot\hat{R})-\hat{\mathcal{P}}_j\cdot\hat{\mathcal{P}}_{j'}\Big)(\frac{\sin\xi}{\xi^3}-\frac{\cos\xi}{\xi^2})\Big],\end{split}\eeq
where $\hat R=(\mathbf r_j-\mathbf r_{j'})/|\mathbf r_j-\mathbf r_{j'}|$ is the displacement unit vector. This function has a well-defined limit as $\xi \rightarrow 0$, $\lim_{\xi\rightarrow 0} f(\xi)=\hat{\mathcal{P}}_{j}\cdot\hat{\mathcal{P}}_{j'}$. For two parallel dipoles, the function is very close to identity in the range of our interest which corresponds to $\xi \ll 1$. Detailed discussions of the function can be found in the literature such as \cite{kiffner, Stephen_JCP1964, Lehm_PRA1970}.

In the resonant dipole-dipole interaction model (RDDI), the donor-acceptor coupling (also termed as the coherent interaction or the RDDI potential) $\Delta$ arises from the vacuum-mediated coupling between the two atoms, and is dependent on the distance between two dipoles. For two parallel dipoles one in the excited state and one in the ground state, $\Delta$ is given as follows \cite{Petro_PRL2002, Lehm_PRA1970, Stephen_JCP1964},
\beq \bsplit
\Delta=\frac{3}{4}\sqrt{\Gamma_{11}\Gamma_{22}} \Big\{&-[1-\cos^2\theta]\frac{\cos\xi}{\xi}\\
+&[1-3\cos^2\theta]\Big(\frac{\sin\xi}{\xi^2}+\frac{\cos\xi}{\xi^3}\Big)\Big\},\label{delta}
\end{split}\eeq
where $\theta$ is the angle between the dipoles and the displacement vector joining the two dipoles and $\xi$ represents the distance between the dipoles measured in terms of cavity wavelength. This expression is valid as long as the energy splitting between the donor and the acceptor is much less than their emission and absorption frequencies and it provides the distance-dependence of the system.

\bigskip
\section{Parameter regimes related to experiments}
The FRET technique is applicable to the nonradiative dipole-dipole coupling where the wavelength of the virtual photon is larger than the distance between the two dipoles. For example, the typical length scale of the donor fluorescence is in the 300-800 nm range, and the FRET in an approximate 1-10 nm range \cite{Medintz_book2014}. This corresponds to $\xi \sim 0.005-0.2$ and the scaling relation $\Delta\sim\xi^{-3}$ holds within the range according to Eqn.~\ref{delta}. When the distance is much less than the length scale, the ideal dipole approximation in F$\ddot{\mathrm{o}}$rster's calculation breaks down. On the other hand, if the distance is much larger than the length scale, the FRET process stops and the intermediate and radiation zones become relevant. The physics discussed in this paper is within the regimes where the FRET technique is applicable.

In the near zone region, the relevant components are those proportional to $\xi^{-3}$ due to the static electric field. Ignore the parts due to the intermediate and far field radiation, and consider the case $\theta=\frac{\pi}{2}$ and $\Gamma_{11}=\Gamma_{22}=\Gamma$, then $\Delta \approx \frac{3\Gamma}{4\xi^3}.$ The coherent interaction $\Delta$ is divergent as $\xi\rightarrow 0$ according to the above definition. When $\xi \rightarrow 0$, the Forster's theory is not applicable and several regularization methods can be performed on $\Delta$ to render it finite \cite{Brooke_PRA2008}. The details of the regularization goes beyond the scope of this study. The typical value of dipole moments  $|\vec{\mathcal{P}}|$ of a strongly dipolar molecule is roughly of order a few debyes. For example, the dipole moment of water molecules is approximately 1.84 debyes, and the chloromethane is about 8.5 debyes. This gives an estimation of the order of magnitude for the dissipation constant $\Gamma_{ii}/\nu\sim 10^{-11}-10^{-9}$. Since $f(\xi)$ is very close to identity function in the range of our interest, we can set the collective decay to be $\Gamma_{ij}=10^{-9}$ in this study, and this correspondingly gives the strength of the coupling $\Delta$. The approximate range of $\Delta$ calculated from Eqn.~\ref{delta} is shown in Fig.~\ref{f_delta}. For the individual decay rate in the quantum master equation of molecules, an additional path from the inhomogeneous broadening that originates from vibrations and other dense medium often contributes crucially to the electronic dephasing. Typically in many dipolar molecules the inhomogeneous broadening $\Gamma_{inh,ii} \sim 100$GHz, which is approximately $10^{-4} \nu$. This effect can lead to a dominant contribution of individual decay rate $\Gamma_{inh,ii}$.  The total individual decay rate $\Gamma_{ii}= \Gamma_{h,ii}+ \Gamma_{inh,ii} \approx \Gamma_{inh,ii}$, where $\Gamma_{h,ii}$ is the decay rate we previously calculated without the subscript ``h'' and $\Gamma_{inh,ii}$ is the additional contribution coming from inhomogeneous broadening.

\section{Resonance energy transfer under strong laser drive}\label{v}
The efficiency of the excitation energy transfer (or quantum yield) $\eta$ is related to the effective energy transfer rate $\kappa$ by $\eta=\frac{\kappa}{\kappa+\Gamma_{11}\rho_{aa}}$ \cite{jianshucao}, and $\eta \approx \frac{\kappa}{\Gamma_{11}\rho_{aa}}$ when the rate is low.  In the nonequilibrium steady state (NESS), the rate for the excitation energy transfer and that of the acceptor's decay to the ground state coincide as the result of the conservation of probability, i.e. $\kappa= \Gamma_{22} \rho_{bb}$. When the intensity of driving field is low, the rate of energy transfer can be achieved perturbatively as a power series of the donor-acceptor coupling $\Delta$, and the leading order in the perturbative expansion is $\Delta^2$. Since $\Delta \sim 1/\xi^3$ where $\xi=d/\lambda_d$ is the distance between two dipoles measured by the laser wavelength, the result reduces to the F$\ddot{\mathrm{o}}$rster regime that the transfer rate against donor-acceptor distance obeys the inverse sixth power scaling \cite{Forster_AP1948}. However, a strong drive field can lead to the efficient energy transfer over long distance between the donor and acceptor molecules, resulting in the breakdown of the perturbative treatment. In what follows, we present the results and demonstrate the possibility of the enhanced long-range energy transfer.

\subsection{Energy transfer in steady state}
The reduced density matrix for NESS can be obtained by solving Eqn.~\ref{master}. The full nonperturbative solution is extremely complicated, but several interesting special cases exist and show the analytical behaviors that are most relevant for our discussion. Our analysis is based on the analysis of the NESS solutions and the dynamics towards the NESS is discussed separately in the next section. 

\subsubsection*{Energy transfer and quantum yield at short distance and near resonance}
\begin{figure}[htb!]
\includegraphics[width=.9\columnwidth]{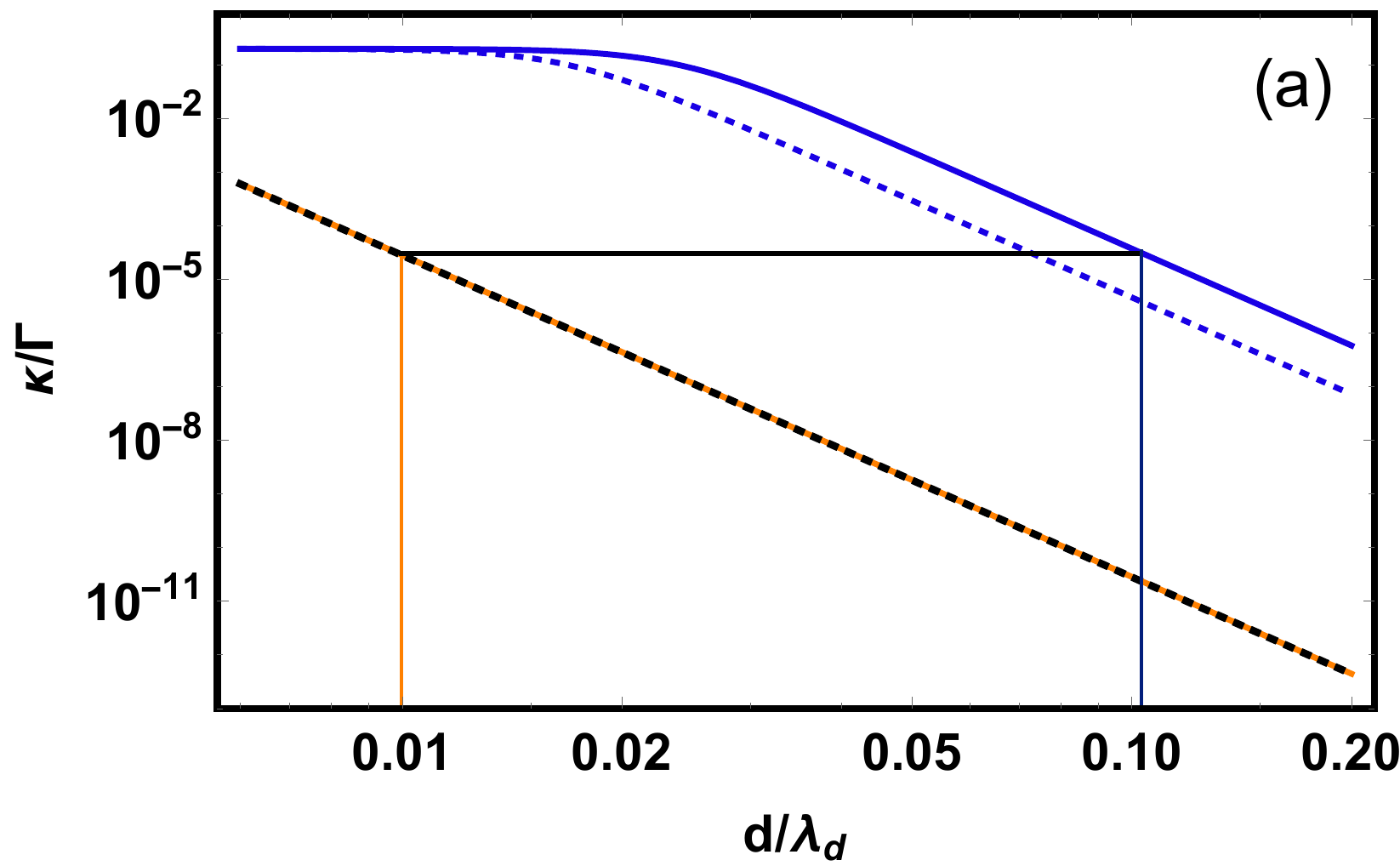}
\includegraphics[width=.9\columnwidth]{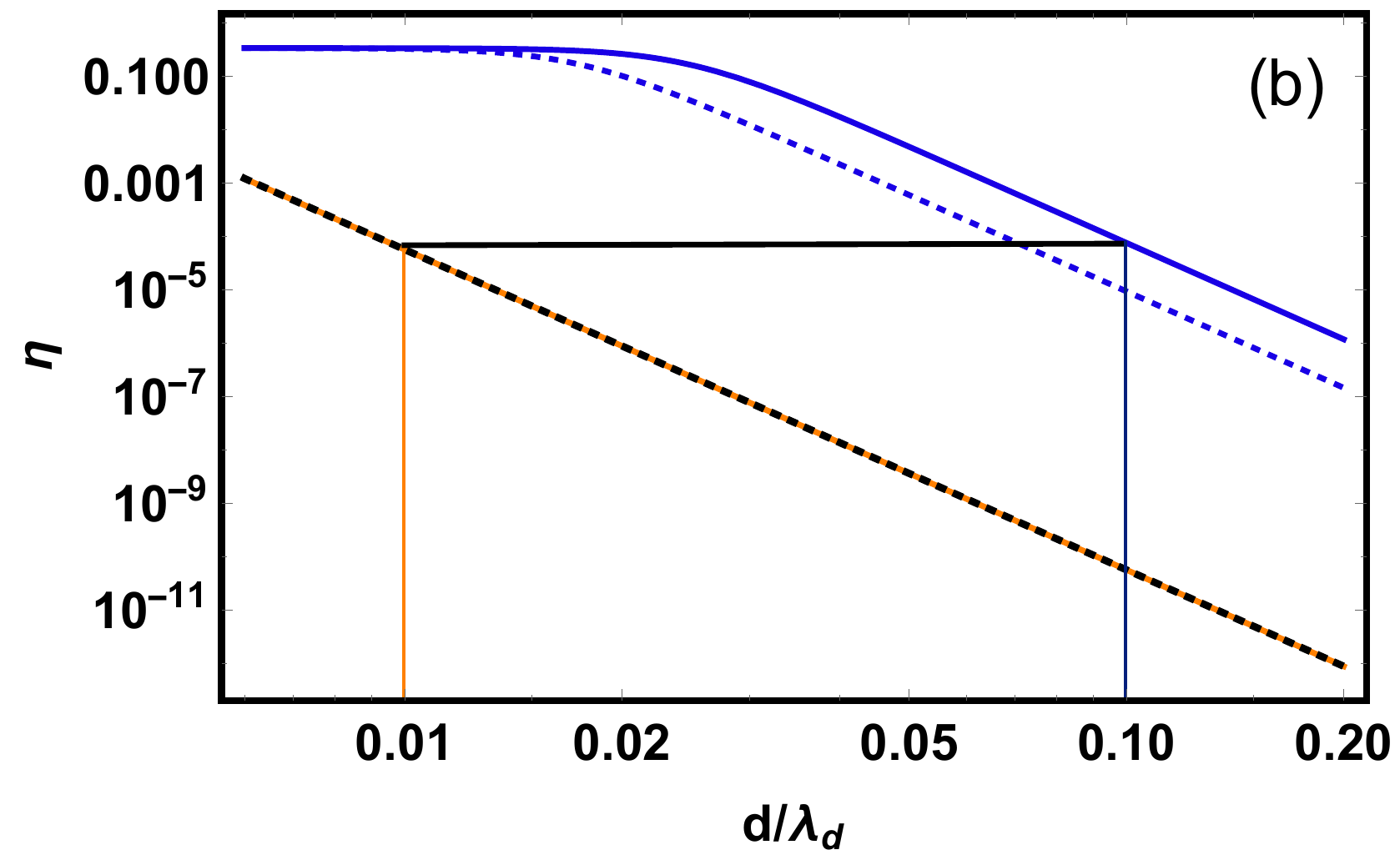}
\includegraphics[width=.9\columnwidth]{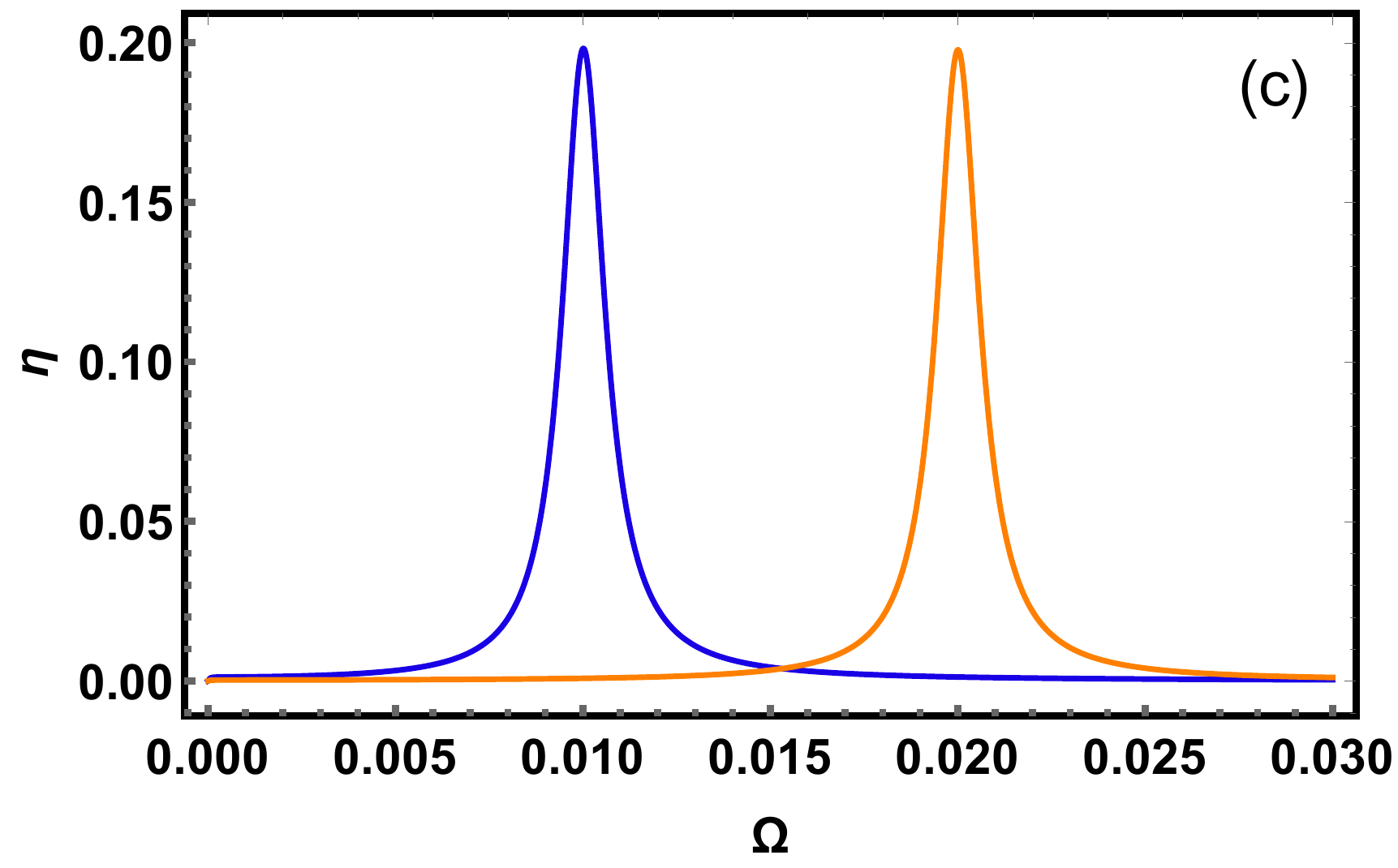}
\caption{(a) Rate of energy transfer vs. donor-acceptor separation $\xi$ in log-log scale. The solid navy line (top): the resonance case ($\Omega=R^*=\omega_1-\omega_2$). The dotted navy line below it: $0.2\%$ off the resonance. The overly high Rabi frequency case ($\Omega=2R^*$) coincides with the rates at low Rabi frequencies $\frac{\Omega}{\omega_1-\omega_2}<0.2$, which are marked by the slanted orange line. For other parameter regimes, the results lie between the navy and the orange line. The dotted black line overlapping with the orange is the weak laser limit (assume $\Omega\gtrsim\Gamma$). $\delta_1=0$, $R^*=0.1$, $\Gamma=10^{-4}$.  (b) Quantum yield vs. distance $\xi$ in log-log scale at the same condition as (a). The efficiency is almost identical to the rate $\kappa$ up to a factor of $\Gamma$ when $\kappa\ll1$.  (c) Quantum yield vs. Rabi frequency at fixed distance. The characteristic Rabi frequencies are set to $R^*=0.01$ (the navy line on the left) and $R^*=0.02$ (the orange line on the right). Here, $\delta_1=0,\ \Delta=5\times 10^{-4},$ and $\Gamma=10^{-4}$.}
    \label{rate}
\end{figure}

In the following, we consider the regime of strong driving field when the Rabi frequency is close to the donor-acceptor energy splitting $\omega_1-\omega_2$. We will refer to this Rabi frequency as the characteristic Rabi frequency $R^*=\omega_1-\omega_2$. For this ``near resonance'' regime, the solution behaves very differently at small distances from the large distance behavior [the navy lines in Fig.~\ref{rate} (a)]. 

For the sake of succinctness, consider zero detuning $\delta_1=\omega_1-\nu = 0$ and symmetric total individual decay constants $\Gamma_{11}=\Gamma_{22}=\Gamma\ll R^*$. In the limit of zero detuning and the Rabi frequency close to the characteristic value $R^*$, the effective energy transfer rate $\kappa$ is approximately given is as follows (the exact solution is convoluted and provided in the appendix),
\beq \bsplit \kappa=\ &\Gamma \left(\frac{1}{5}-\frac{3}{50} \gamma-\frac{4}{75}(\frac{\Delta}{R^*})\gamma\ \tilde\Omega+O\left(\gamma^2\right)+O(\tilde\Omega^2)\right)\\
\approx &\ \frac{\Gamma}{5}-\left(\frac{8}{75}-\frac{64}{675}(\frac{\Omega-R^*}{R^*}) \right)\frac{\Gamma^3}{\Gamma_h^2} \xi^6\,,
\end{split} \label{kappar}
\eeq
where $\gamma=\frac{\Gamma^2}{\Delta^2}\approx \frac{16}{9}\frac{\Gamma^2}{\Gamma_h^2}\xi^6$ is a small parameter at short distances, and $ \tilde\Omega=\frac{\Omega-R^*}{\Delta} $ measures the bias between the Rabi frequency and the donor-acceptor level splitting in the scale of the dipole-dipole potential. We will refer to Eqn.~\ref{kappar} as the ``strong coupling expansion'' as it is accurate at short distances when the coupling between the molecules is strong.

\afterpage{\FloatBarrier} 

The leading order term predicts that the transfer rate stays nearly constant in short distances. Eqn.~\ref{kappar} shows the super energy transfer such that the constant term independent of donor-acceptor distance dominates in the vicinity of $R^*$, the energy splitting, where the perturbative analysis breaks down. This leading term shows that that the optimal the transfer rate is approximately $\frac{\Gamma}{5}$. This is in accordance with Fig.~\ref{rate} (c) where the peaks are about $\frac{1}{5}$. The range of effectiveness for the distance-independent transfer rate extends until the distance $\xi \gtrsim \min(\sqrt[3]{\frac{3\Gamma_h}{4|\Omega-R^*|}},\sqrt[3]{\frac{3\Gamma_h}{4\Gamma}})$. It subsequently begins to decay, roughly in the manner of $(1-\frac{8\Gamma^2}{75\Gamma_h^2}\xi^{6})$ [Fig.~\ref{rate} (a, b) navy lines]. This near-constant decay mode with distance allows the energy transfer to be effective in a larger range.

Fig.~\ref{rate} (a,b) shows the rate and the efficiency of the excitation energy transfer against the distance between the donor and the acceptor. The near resonance cases are the top navy curves and the off-resonance results are the dotted and orange lines. Physics at short distances (the leveling off and the curving down segment) of the navy curves in Fig.~\ref{rate} (a,b) is described by the strong coupling formula Eqn.~\ref{kappar} and the segment that drops at $\xi^{-6}$ (the slanted straight lines) can be described by the weak coupling expansion Eqn.~\ref{kappa} introduced below. The vertical lines intersecting with the horizontal axis indicate the distances at which the transfer rates and quantum yield of the two regimes (in and off resonance) drop to certain detectable values, and the black horizontal line connecting the in-resonance curve on top and the off-resonance curve measures the distance increased by the non-perturbative phenomenon. The distance of effective energy transfer is increased by approximately \textit{one order of magnitude} through applying the strong driving field at the resonant Rabi frequency. This increases the effective FRET from $\sim$ 10nm to $\sim$100nm range. Assuming that the Rabi frequency of the laser-system, which is tuned to the energy splitting of the system, has an error $\sigma$, the rate/efficiency of the excitation energy transfer will begin the rapid drop of $\xi^{-6}$ at a shorter distance $\xi^*\approx \sqrt[3]{3\Gamma/(4 \sigma)}$, as is indicated in the dotted navy lines in Fig.~\ref{rate} (a, b)].

At large distances ($\xi \gg \min(\sqrt[3]{\frac{3\Gamma_h}{4|\Omega-R^*|}},\sqrt[3]{\frac{3\Gamma_h}{4\Gamma}})$), $\gamma$ and $\tilde\Omega$ are of order one or above, the strong coupling expansion Eqn.~\ref{kappar} breaks down and the exact result approaches the weak coupling expansion Eqn.~\ref{kappa}. To be more specific, the strong coupling expansion above is accountable for the near-resonance phenomenon at short distances, while the weak coupling expansion to be introduced later describes the physics in the complementary regimes, i.e. at large distances or off resonance. At short distances, the rate of energy transfer from the off-resonance mechanism [Eqn.~\ref{kappa}] is suppressed roughly by a factor of $~\tilde\Delta^2$ compared to that of in resonance mechanism Eqn.~\ref{kappar}. The efficiency of the excitation energy transfer $\eta=\frac{\kappa}{\kappa+\Gamma_{11}\rho_{aa}}\sim\kappa$ is approximately proportional to the rate as the result of low transfer rates and slowly changing donor population [Fig.~\ref{rate} (b)].

\subsubsection*{Energy transfer and quantum yield at large distance or off resonance}
In the regimes where the above approximation becomes invalid, namely at large distances or off resonance,  we refer to the following expansion to describe the physics occurred. Under the assumption that $\Omega\gg\Gamma \gg \Gamma_{12}$ and in the vicinity of zero detuning $\nu=\omega_1$, the leading order of $\kappa$ in $\tilde\Delta$ and $\Gamma$ is given as follows,
\bea  \label{kappa}
\kappa&=& \Gamma \cdot  \frac{2\Omega ^2+R^{*2}+O\left(\Gamma^2\right)}{2R^{*2}}\tilde\Delta^2+O\left(\tilde\Delta^4\right)\\
& \approx & \frac{9\left(2\Omega ^2+R^{*2}\right)\Gamma \Gamma_h^2}{32\left(R^{*2}-\Omega^2\right)^2}\cdot \xi^{-6}\,,
\eea
where 
\beq \tilde\Delta^2=\frac{\Delta^2}{\left(R^{*2}-\Omega^2\right)^2/R^{*2}+O(\Gamma^2)} \eeq
is the dimensionless coupling. We will refer to Eqn.~\ref{kappa} as the ``weak coupling expansion'' as it describes the weak coupling scenario and the long distance behavior. The coupling in the case of nonradiative energy transfer is give by $\Delta=\frac{3\Gamma_h}{4\xi^3}$. We notice that in the off-resonance, weak laser regimes, i.e. $\Omega\ll\omega_1-\omega_2=R^*$, the rate of transfer returns to that of the perturbative result 
\beq \label{kappaxi}
\kappa\approx\Gamma \cdot \frac{\Delta^2}{2R^{*2}}\approx \frac{9\Gamma \Gamma_h^2}{32R^{*2}\xi^6},
\eeq
which is independent of the Rabi frequency. This case returns to the F\"orster's scaling in the FRET regime when the Rabi frequency is sufficiently below the energy splitting between the two molecules and we will not discuss further. We shall note that for extremely large Rabi frequencies ($\Omega>R^*$), though this expansion is also accurate but the rate of transfer then is dependent on the Rabi frequency and decreases as the driving field is strengthened [Fig.~\ref{rate} (c)].

The above weak coupling expansion should not be understood as the perturbative result. It is the non-perturbative formula and it recovers the perturbative result in the low Rabi frequency limit. The approximation in the premise is applicable when $\tilde \Delta^2 \ll 1$. In another word, we should refer to it when either $\Delta^2\ll\Gamma^2$ or $\Delta^2\ll \frac{(R^{*2}-\Omega^2)^2}{R^{*2}}$ is satisfied. The first condition corresponds to $\xi\gg\sqrt[3]{\frac{3\Gamma_h}{4\Gamma}}$, which is the long distance condition complementary to the strong coupling regime. The second condition is equivalent to $|\Omega-R^*|\gg\Delta$, which is the off-resonance condition that requires the energy difference between the donor and the acceptor away from the Rabi frequency. 

In Fig.~\ref{rate} (a, b) the black horizontal lines give the scale of the distance the enhanced energy transfer enlarges before it drops to the same level as that derived from the FRET mechanism. The slanted orange lines and the overlapping dotted lines are for the off-resonance regimes. The orange lines represent two different cases of the driving field of which the quantum yields are indistinguishable---one is for the low intensity laser ($\Gamma<\Omega<20\%R^*$) and the other is for the extremely strong laser ($\Omega=2R^*$). For the Rabi frequencies below $20\%$ of the energy splitting $R^*$, the differences in the rates of energy transfer are indistinguishable. This line provides the scale of the transfer rates in the off-resonance perturbative regime. For the Rabi frequencies between the two extreme limits, the results lie between the solid navy line and the orange line. The dotted line that overlaps with the orange is the leading order approximation given by Eqn.~\ref{kappaxi}. 

The above results derived from the quantum master equation approach show, for the in-resonance regime, a slower decay of the energy transfer than that of the inverse sixth power when the donor and the acceptor are put in short distances. The boost in the rate and the efficiency of the energy transfer are significantly more prominent when the acceptor is placed further away from the donor due to the different scalings at short and large distances, which allows the transmission to be effective across a larger range. A non-linearity of the energy transfer with the strength of the laser is noticed, with the peak at $\Omega=R^*$ for fixed distances.

\begin{figure}[ht]
\centering
{\includegraphics[width=.9\columnwidth]{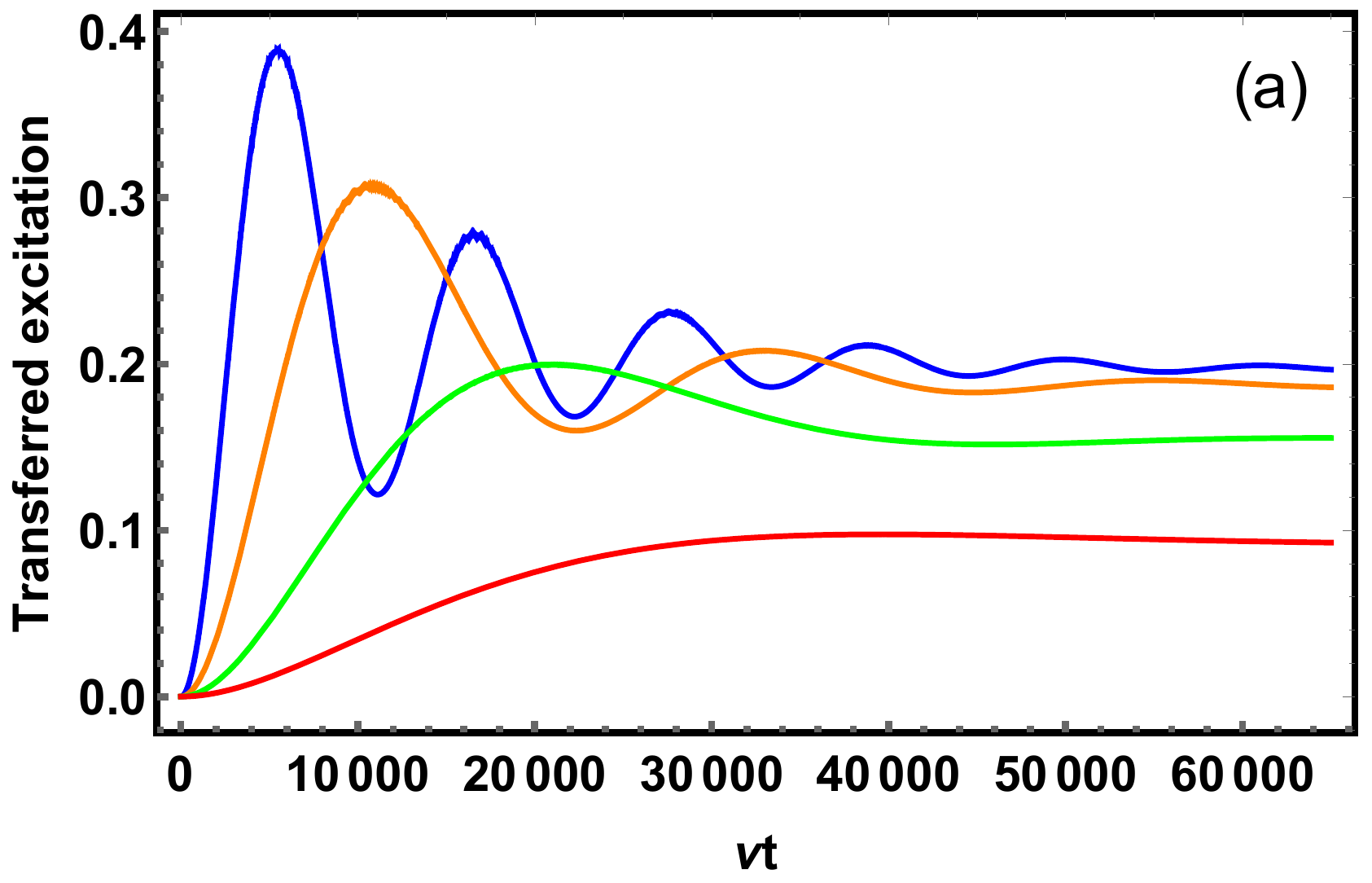}}
{\includegraphics[width=.9\columnwidth]{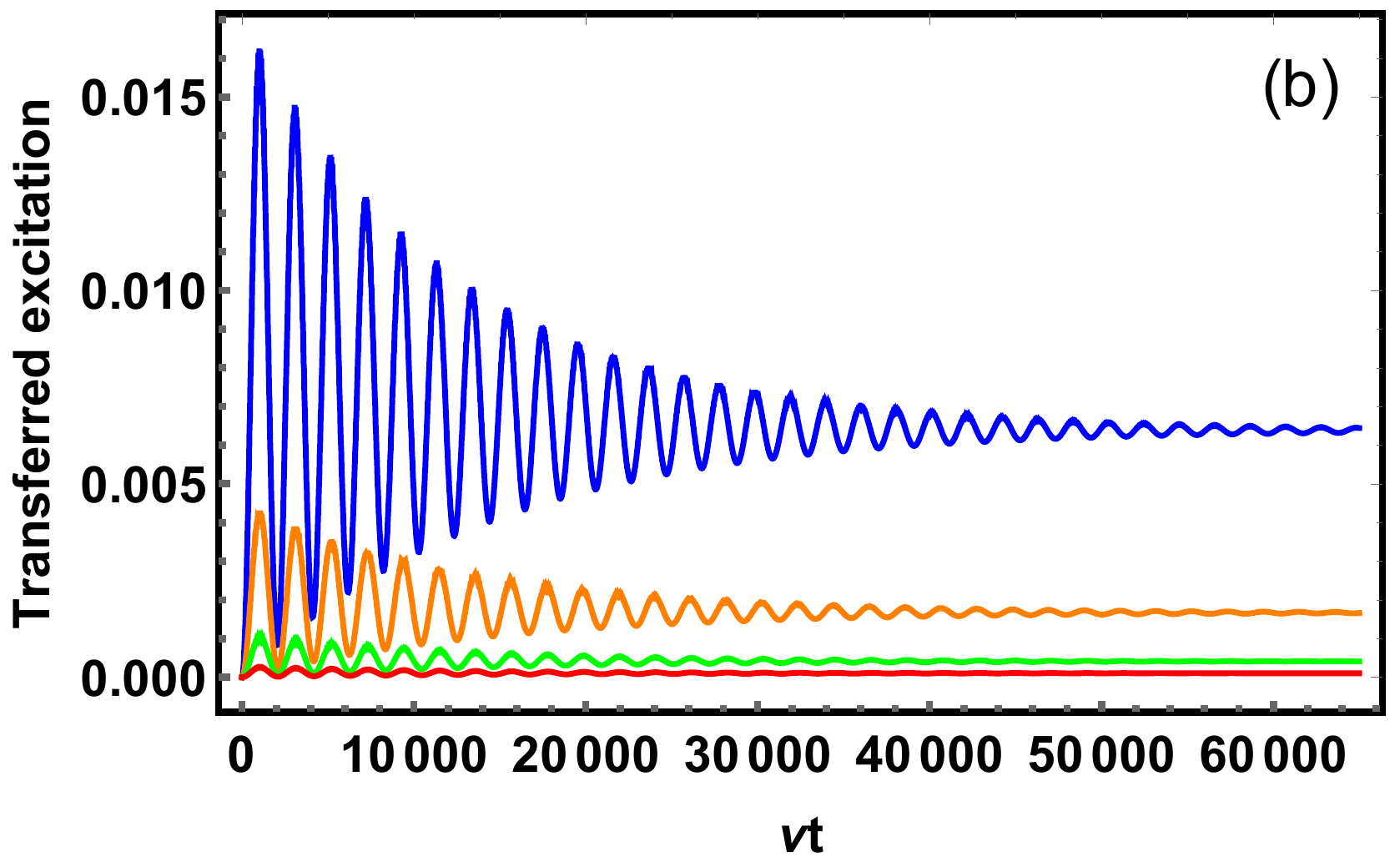}}
\caption{The dynamics of the excitations on the acceptor state with the time multiplied by laser frequency $\nu$. (a) In resonance $\Omega=R^*$. (b) Off resonance $\Omega=R^*(1\pm 3\%)$. The color code is given as follows (from top to bottom), the blue lines: $\Delta=4\Gamma$, the orange lines: $\Delta=2\Gamma$, the green lines: $\Delta=\Gamma$, and the red lines: $\Delta=0.5\  \Gamma$. For (a) and (b), $\Gamma=10^{-4}$, $\Gamma_h=10^{-9}$ and $\omega_1-\omega_2=0.1$.}
    \label{dy}
\end{figure}

\subsection{Dynamical relaxation towards NESS}
The above discussions are based on the NESS solutions. The validity of the NESS solution can be verified by examining the eigenvalues of the Liouville superoperator. We study the dynamics of the system initially prepared to be on the ground state and irradiated by the laser beam at time $t=0$. 

For the resonance case $\Omega=\omega_1-\omega_2$, the dynamics can be roughly divided into three regimes, the under-damped regime ($\Delta >\Gamma$), the critically-damped regime ($\Delta=\Gamma$), and the over-damped regimes $(\Delta<\Gamma)$. In the under-damped regime, the system oscillates around the value of the NESS and the amplitude of oscillation decays with time [Fig.~\ref{dy} (a)]. For the parameters used in Fig.~\ref{dy} (a), the NESS can be approximated by the strong coupling expansion Eqn.~\ref{kappar} where the leading term is dominated by the constant $\frac{1}{5} \Gamma$ plus a negative correction. Therefore, the asymptotes of the four lines are all close to or slightly less than $\frac{1}{5}$. In the over-damped case, the oscillation is replaced by a steadily-increasing curve as the excitation number in the acceptor accumulates with time. On the other hand, examining the dynamics of system at $\Omega\ne\omega_1-\omega_2$, it can be easily noticed that the asymptotic values for the different couplings are orders-of-magnitude apart, and that the system oscillates at a dramatically higher frequency than the resonance scenario without the occurrence of the critically- and over-damped phases as in the in-resonance scenarios. The rationale is as follows. Since asymptotic behavior for systems with parameters given in Fig.~\ref{dy} (b) can be roughly deduced from the weak-coupling Eqn.~\ref{kappa}, reducing $\Delta$ by a factor of two results in a factor four folding in the transfer rate $\kappa$, which is in accordance with Fig.~\ref{dy} (b). Meanwhile, the condition that the energy splitting $\omega_1-\omega_2\gg\Delta$ ensures that even a minuscule percentages of deviation of the Rabi frequency from $R^*$ will cause the frequency difference between the dressed donor state and the acceptor state larger than the decay rate $\Gamma$, which renders the system highly oscillatory.

\subsection{Optimal quantum yield and Rabi splitting}\label{note}

\begin{figure}[htbp!]
\centering
\includegraphics[width=.9\columnwidth]{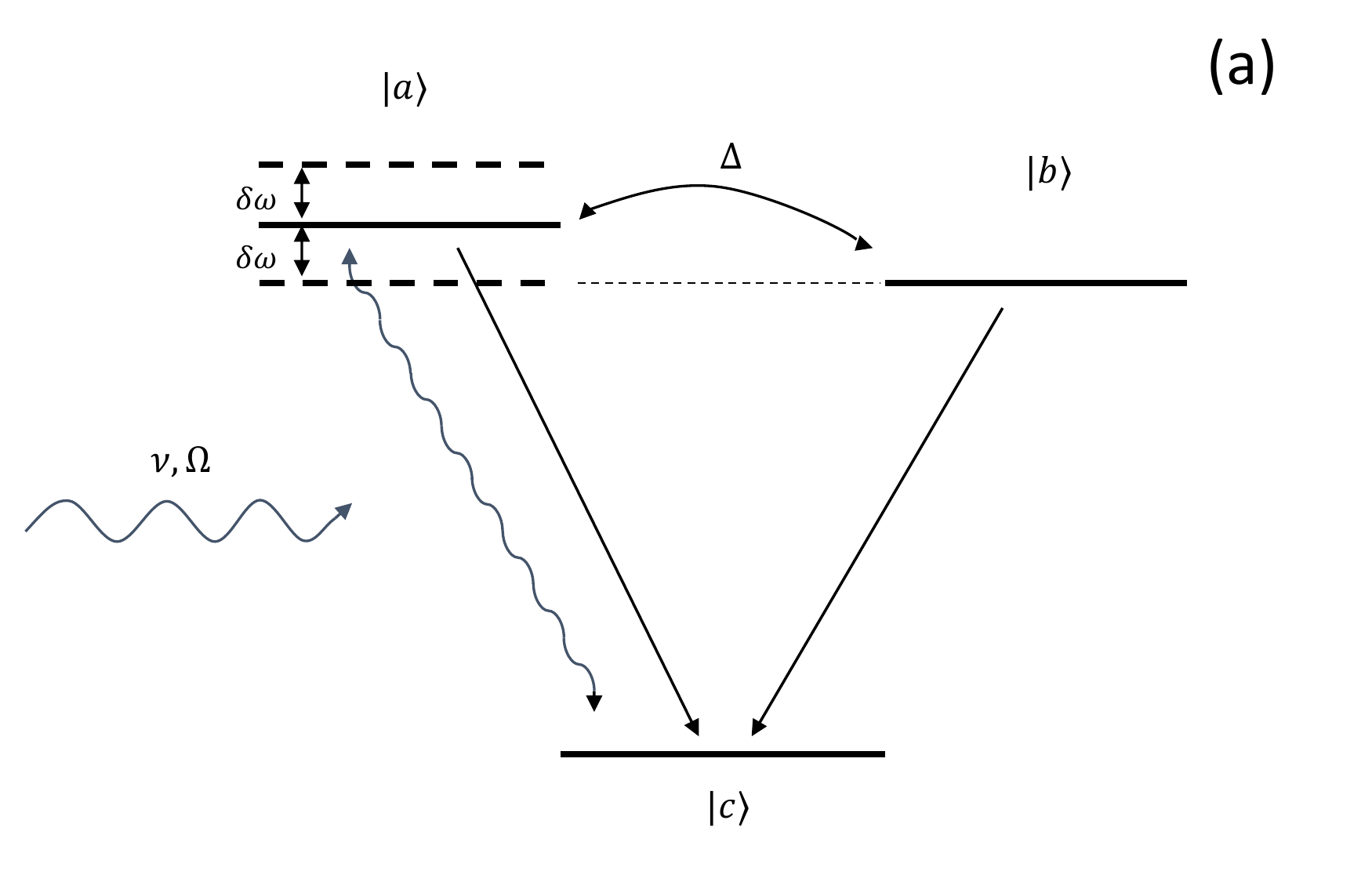}
{\includegraphics[width=.9\columnwidth]{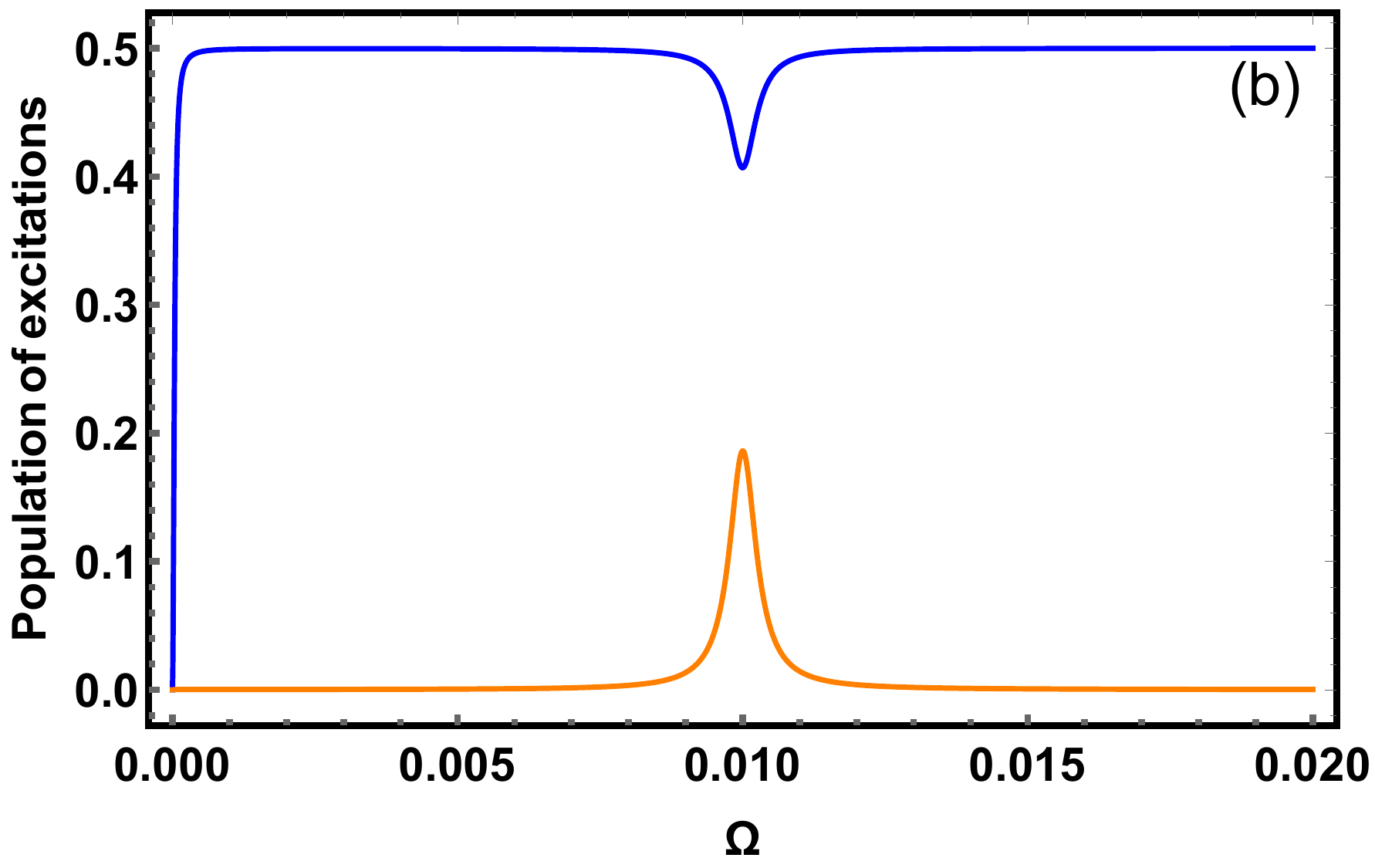}}
{\includegraphics[width=.9\columnwidth]{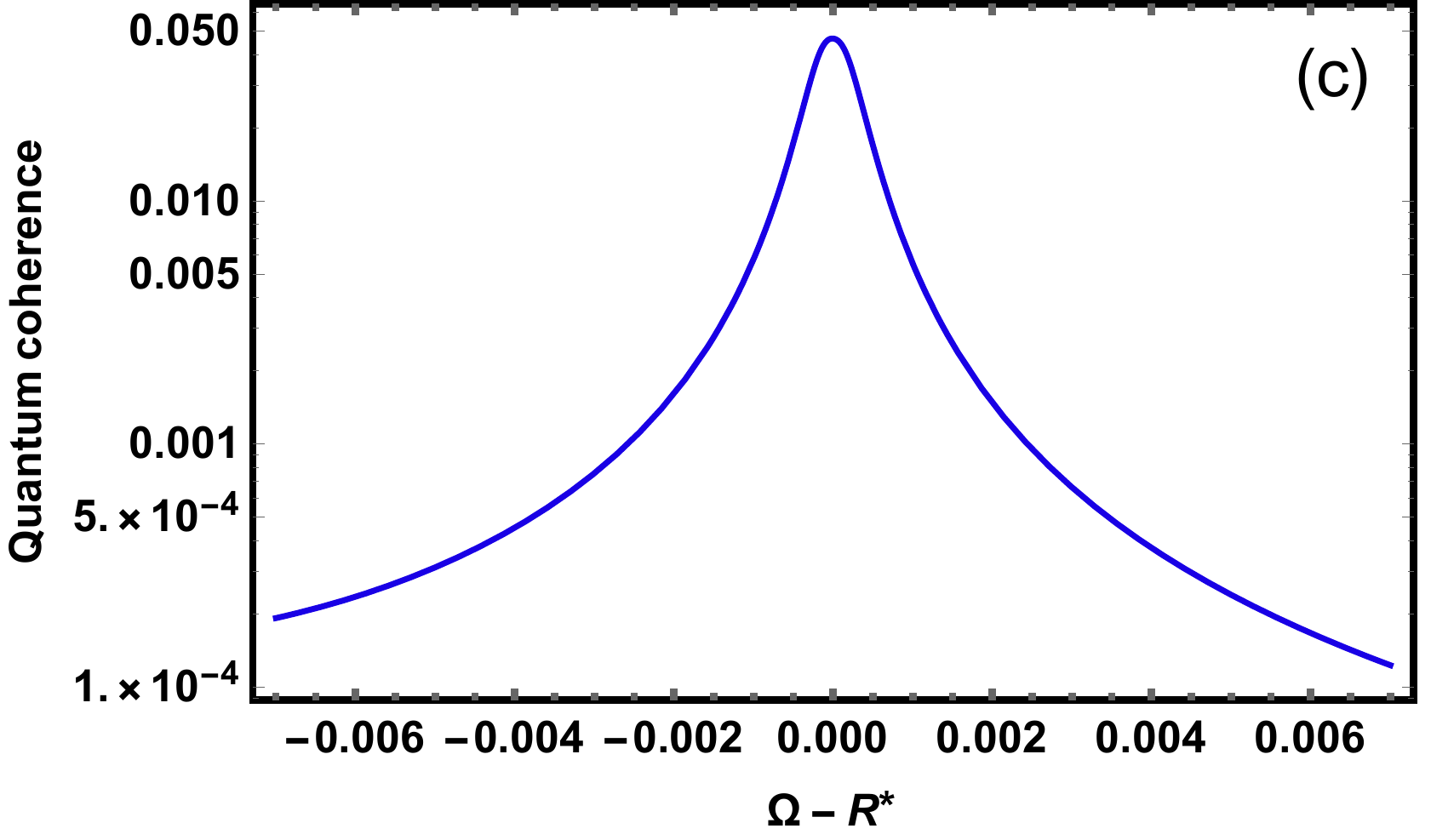}}
\caption{(a) Schematic demonstration the energy level splitting due to the dynamic Stark effect. (b) Population on state ``a'' (navy, top) and ``b'' (orange, bottom) vs Rabi frequency $\Omega$. $\delta_1=0,\ R^*=0.01.$ The peak value of excitation of around 0.2 can be inferred from the constant term in Eqn.~\ref{kappar}. (c) Quantum coherence vs bias of Rabi frequency from the $R^*$ in logarithmic scale. The off diagonal element of the density matrix corresponding to the coherence between the excitation on the donor and the excitation on the acceptor. For all, $\Delta=2\times 10^{-4}$, $\Gamma=10^{-4}$.}
    \label{2}
\end{figure}

In the weak laser limit, the laser beam has limited impact on the system besides the excitation of the irradiated molecule, and the process of energy transfer can be calculated assuming no change in the energy levels of the system. However, under a strong laser drive, the molecule-photon interaction mixes the electron and photon degrees of freedom, resulting in the Rabi splitting or dynamic stark splitting. This is the reason for the non-linearity in the efficiency of the energy transfer with the laser intensity and it happens when the Rabi frequency of the laser-donor system is close to the energy bias between the donor and the acceptor molecules. In the vicinity of this regime, the perturbative treatment of the energy hopping needs to be modified to incorporate all the non-perturbative and quantum effects.

When the donor molecule is irradiated by the driving field with frequency close to its excitation energy, the dressed state of the donor splits into effective levels different from the original level [see Fig.~\ref{2} (a)]. In other words, the state ``a'' splits into two states with effective frequencies $\omega_1 \pm \Omega$. Naively, as the laser intensity is strengthened we expect an increase in the acceptor excitations due to the boost of the driving field. This is true when the laser strength is weak with Rabi frequencies less than the energy splitting. At sufficiently large Rabi frequencies, one effective level of the dressed donor state ``a'' $\omega_1 \pm \Omega$ can drop below the level of the acceptor if further strengthen the intensity of the drive. When $\Omega=\omega_1-\omega_2$, the lower dressed state is in resonance with the acceptor energy level, leading to the maximal population at the acceptor and the energy transfer is optimized. The peak value of excitation on ``b'' around 0.2 can be inferred from the constant term in Eqn.~\ref{kappar}. The more general case that includes the detuning between the cavity frequency and the donor energy level is not the main focus of the study and is discussed in the Appendix. 

Besides the excitation of the donor states and the Rabi splitting, the applied strong laser drive also has the effect of magnifying the quantum coherence between the donor and the acceptor. This quantum effect has the most obvious manifestation in the vicinity of the characteristic Rabi frequency $\Omega=R^*$ [see Fig.~\ref{2} (c)]. These effects near the characteristic Rabi frequency render the traditional analysis inaccurate.

\section{Conclusion}
We presented a non-perturbative approach for the study of the excitation energy transfer in molecular systems under a strong drive field. The great enhancement of the long-range energy transfer between donor and acceptor has been observed when the system is strongly driven. Our results show that the optimal transfer appears at a certain driving power instead of increasing co-linearly along with the strengthening of the laser, which results from the dynamic Stark effect. Notably, when the laser intensity is modulated, the rate and the efficiency of the energy transfer all reach the maximal values at a characteristic Rabi frequency holding the distance between the donor and the acceptor fixed. However, a significantly larger improvement is witnessed at large distances, which allows an extended range of effective energy transfer compared with the case when the applied Rabi frequency is off the particular strength. This is because at short distances the scaling of the energy transfer follows a different pattern, which allows the efficiency and the rate to be roughly constant as the donor and acceptor are pulled apart. In most cases away from the characteristic Rabi frequency, we found that the general $R^{-6}$ scaling is an accurate approximation even when a strong laser is applied. The different distance scaling starts to appear when the strength of the laser is tuned close to the characteristic value, resulting in a much slower decay than the $R^{-6}$ scaling from the F\"orster mechanism. The range of the effective molecular energy transfer is enhanced by at least one order of magnitude near this characteristic value compared with that when the Rabi frequency is off. In the limit of weak driving or large bias from the Rabi frequency, the discussion reproduces the results given by the classical F\"orster's theory. The longer-range and enhanced energy transfer demonstrated by our model can be further measured by the time-resolved fluorescence spectroscopy. In light of the recent advance in ultrafast nonlinear spectroscopy for nano-scaled materials, our work offers insights and perspectives to the spectroscopic study of cooperative energy transfer in molecular polaritons involving a large number of molecules.

\begin{acknowledgments}
X.W thanks Kun Zhang for the comments on the early version of the manuscript.  Z.D.Z gratefully acknowledges the support of ARPC-CityU new research initiative/infrastructure support from central (No. 9610505).
\end{acknowledgments}

\begin{widetext}
\section*{Appendix}
\begin{figure}[H]
\centering
\includegraphics[width=.45\columnwidth]{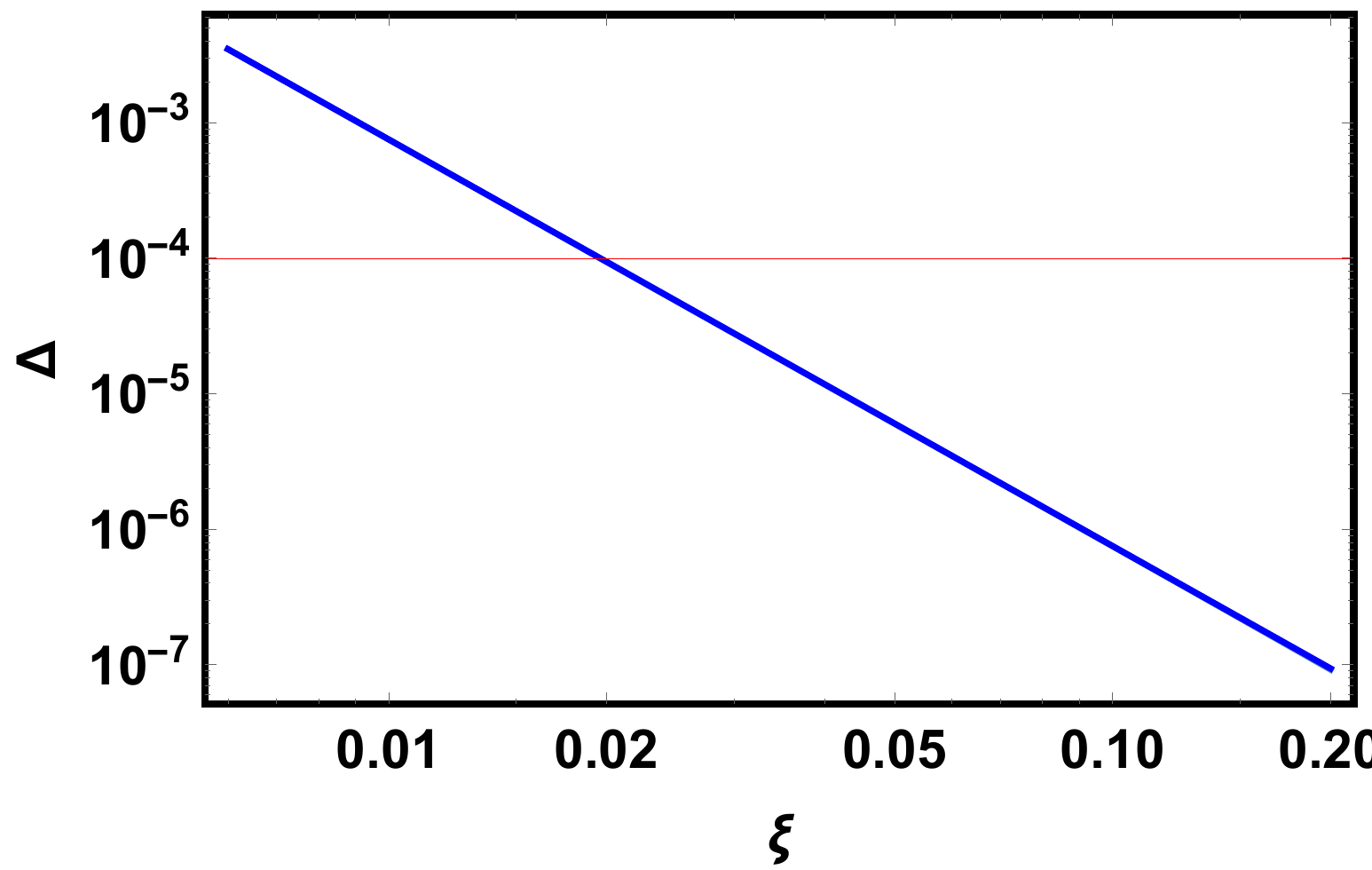}
\caption{The typical range of the coupling potential $\Delta$ in a FRET process as a function of inter-molecular distances $\xi$ in log-log scale. The horizontal line in the middle is the value of $\Gamma=10^{-4}$. Here, we assume $\Gamma_h=10^{-9}$.}
    \label{f_delta}
\end{figure}

\subsection*{Equation of motion for the system}
We redefine the system Hamiltonian as
\beq h_s \equiv \nu(|1\rangle\langle 1|+|2\rangle\langle 2|)\,,
\eeq
and rewrite the total Hamiltonian, $\calh=(h_s+\calh_R)+(\calh_S-h_s+\calh_I) \equiv h_0+\calh_{int}$, where $h_0=h_s+\calh_R$ is treated as the free Hamiltonian and $\calh_{int}=\calh_S-h_s+\calh_I$ is the new interaction term.  With the free Hamiltonian and the interaction component defined above, the interaction Hamiltonian $\calh_{int}$ in the interaction picture takes the form $ \tilde \calh_{int}(t)= e^{ih_0 t}(\calh_S-h_s+\calh_I)e^{-ih_0 t}\equiv \til \calh_0(t) +\til \calh_I(t)$, where $\til\calh_0(t)= e^{ih_0 t}(\calh_S-h_s)e^{-ih_0 t}$ and $\til \calh_I(t)=e^{ih_0 t}\calh_I e^{-ih_0 t}$. 

Now that $\til \calh_0(t)$ is time independent, it can be calculated explicitly using the Baker-Hausdorff formula,
\beq \til\calh_0=\sum_{j=1}^2 \delta_j |j\rangle\langle j|+\big(\Delta |1\rangle \langle 2|+\Omega|1 \rangle \langle 0|+h.c. \big)
\eeq
where $\delta_j=\omega_j-\nu$ is the energy difference between the laser and the $i^{\rm{th}}$ excited state. Similarly, the interaction Hamiltonian is 
\beq 
\calh_{I}=\ -\sum_{j=1}^2\sum_{s}\lambda_s^{(j)}v_s^{(1)}(b_s^{\left(j\right)}+b_s^{\left(j\right),\dag})|j\rangle \langle j| +\sum_{j=1}^2\sum_{k,\sigma} g^{j}_{k,\sig}|j\rangle \langle 0|a_{k,\sig}+ g^{*j}_{k,\sig}|0 \rangle \langle j|a^\dg_{k,\sig}\,,
\eeq

The equation of motion (von-Neumann equation) for the density operator is determined by $\dot {\til \rho}(t)=i \left[ \til \rho(t),\til \calh_{int}(t) \right] =i \left[ \til \rho(t),\til \calh_0\right] +i\left[\til\rho(t), \til \calh_I(t) \right]$. Inserting the formal solution $\til \rho(t)=\til \rho(0)+i\int_0^{t} ds [\til\rho(s),\til\calh_{int}(s)]$ only back to the interaction term $i\left[\til\rho(t), \til \calh_I(t) \right]$ gives the equation of motion for the total density operator,
\beq \bsplit
\dot{\til\rho}(t)&=i \left[ \til \rho(t),\til \calh_0\right] \\ &+ i \Big[ \til \rho(0)+i\int_0^{t} ds [\til\rho(s),\til\calh_{int}(s)],\til \calh_I(t) \Big].
\end{split} \eeq

We apply the Born-Markov approximation and the Weisskopf-Wigner approximation, ignoring the Cauchy principal value of the integral $\int_0^\infty d\tau e^{ix\tau}=\pi\delta(x)+i\mathcal{P}_c1/x$ and tracing out the bath and the vibrational degrees of freedom, then we have the reduced equation of motion of the system,
\beq \bsplit
\dot{\til\rho}_s(t)= i \left[ \til \rho_s(t),\til \calh_0\right]+\Big( \sum_{j,j'}(1+ n_\nu)\frac{\Gamma_{jj'}}{2}\big(|0\rangle \langle j|\til\rho|j'\rangle \langle 0|-\til \rho |j \rangle \langle j'|\big)+\sum_{j,j'} n_\nu \frac{\Gamma_{jj'}}{2} \big(|j\rangle \langle 0|\til\rho|0\rangle \langle j'| -\delta_{jj'}\til\rho |0\rangle \langle 0|\big) \\
+\sum_{j}{\frac{\Gamma_{inh,jj}}{2}\left(\left|\left.j\right\rangle\left\langle j\right.\right|{\widetilde{\rho}}_s\left|\left.j\right\rangle\left\langle j\right.\right|-{\widetilde{\rho}}_s\left|\left.j\right\rangle\left\langle j\right.\right|\right)+h.c.\Big)}
\label{fullmaster}
\end{split} \eeq
where $n_\nu$ is the number of particles at frequency $\nu$ in the environment which is set to be zero for the consideration of vacuum surroundings, $\Gamma_{jj'}$ is the dissipation constant which is dependent on the distance between the two molecules, and $\Gamma_{inh,jj}$ is the inhomogeneous broadening from the vibrations of the molecules.

\subsection*{The steady state solution}
In Section \ref{v} we discussed the small parameter expansions of the nonperturbative solutions of the NESS. In general, the steady state solution can always be achieved by solving the equation of motion Eqn.~\ref{fullmaster}, $\dot{\til\rho}_s(t)=0$. Solving for the solution of the most general scenario, nevertheless, can be extremely involved and it can be simplified after making the assumption that $\Gamma_{11}=\Gamma_{22}=\Gamma$. To simply our discussion, we present the solution under the above assumption which is assumed in all of the calculations in the paper.

We can reformulate the equation of motion in the Liouville space, where the system density matrix can be expressed in terms of a nine-tuple,
\begin{equation}
    |\rho_s\rangle=\left(\rho_{11},\ \rho_{22},\ \rho_{33},\ \Re(\rho_{01}),\ \Im(\rho_{01}),\ \Re\left(\rho_{02}\right),\ \Im(\rho_{02}),\ \Re(\rho_{12}),\ \Im\left(\rho_{12}\right)\right)^T
\end{equation}
and $T$ represents the matrix transpose. In the Liouville space, the equation of motion takes the following form,
\beq
|\dot\rho_s\rangle=\mathcal M |\rho_s\rangle\,.
\eeq
The matrix representation of the time evolution operator $\mathcal M$ in the Liouville space can be derived from the density matrix equation [Eqn.~\ref{master}].  Under the assumption of $\Gamma_{11}=\Gamma_{22}=\Gamma$, $\mathcal{M}$ is given as follows,
 \beq
\mathcal{M}=
\left(
\begin{array}{ccccccccc}
 -2 n_{\nu } \Gamma  & \left(n_{\nu }+1\right) \Gamma  & \left(n_{\nu }+1\right) \Gamma  & 0 & -2 \Omega  & 0 & 0 & 2 \left(n_{\nu }+1\right) \Gamma _h & 0 \\
 n_{\nu } \Gamma  & -\left(n_{\nu }+1\right) \Gamma  & 0 & 0 & 2 \Omega  & 0 & 0 & -\left(n_{\nu }+1\right) \Gamma _h & -2 \Delta  \\
 n_{\nu } \Gamma  & 0 & -\left(n_{\nu }+1\right) \Gamma  & 0 & 0 & 0 & 0 & -\left(n_{\nu }+1\right) \Gamma _h & 2 \Delta  \\
 0 & 0 & 0 & -\frac{3n_{\nu }+1}{2} \Gamma  & -\delta _1 & -\frac{n_{\nu }+1}{2} \Gamma _h & -\Delta  & 0 & 0 \\
 \Omega  & -\Omega  & 0 & \delta _1 &  -\frac{3n_{\nu }+1}{2} \Gamma   & \Delta  & -\frac{n_{\nu }+1}{2} \Gamma _h & 0 & 0 \\
 0 & 0 & 0 & -\frac{n_{\nu }+1}{2} \Gamma _h & -\Delta  & -\frac{3n_{\nu }+1}{2} \Gamma  & -\delta _2 & 0 & \Omega  \\
 0 & 0 & 0 & \Delta  & -\frac{n_{\nu }+1}{2} \Gamma _h & \delta _2 &-\frac{3n_{\nu }+1}{2}\Gamma  & -\Omega  & 0 \\
 n_{\nu } \Gamma  & -\frac{n_{\nu }+1}{2} \Gamma _h & -\frac{(n_{\nu }+1}{2} \Gamma _h & 0 & 0 & 0 & \Omega  & -\left(n_{\nu }+1\right) \Gamma  & \delta _1-\delta _2 \\
 0 & \Delta  & -\Delta  & 0 & 0 & -\Omega  & 0 & \delta _2-\delta _1 & -\left(n_{\nu }+1\right) \Gamma  \\
\end{array}
\right)\,,
\eeq
where $\Gamma_h$ is the cross decay constant, and $n_{\nu}$ is the photon density at frequency $\nu$.

We are interested in the regime of the steady state solution where the ambient temperature is low compared to the energy levels of the molecules, and thus the thermal excitation is rare. This mimics the experimental situation where the ambient photon bath acts like vacuum---only spontaneous decay is considered and the thermal excitation from the photon bath is absent. To simplify the calculation, we divide the system density operator into two parts: $|\rho_s\rangle=\left(\vec\rho_p,\vec\rho_c\right)^T$ with population terms $\vec\rho_p$ (diagonal terms) and coherence terms $\vec\rho_c$ (off-diagonal terms). Similarly, the time evolution matrix $\mathcal M$ can be expressed in the block form:
\beq \mathcal{M}=
\left(\begin{array}{cc}
  \mathcal{M}_{pp}	& \mathcal{M}_{pc} \\
  \mathcal{M}_{cp}	& \mathcal{M}_{cc}
\end{array}\right)
\eeq
The steady state of the system corresponds to $\mathcal{M}|\rho^{\text{ss}}_s\rangle=0 $, where the superscript ``ss'' stands for the ``steady state''. We substitute the coherent term $\rho_c=-\mathcal{M}_{cc}^{-1}\mathcal{M}_{cp} \rho_p$ into the equation, and obtain a linear equation for the population terms in steady state $\mathcal{A}|\rho^{\text{ss}}_p\rangle=0$, where $\mathcal{A}$ is 
\beq \mathcal{A}= \mathcal{M}_{pp}-\mathcal{M}_{pc}\mathcal{M}_{cc}^{-1}\mathcal{M}_{cp}. \eeq
Once obtaining the steady state solution for $|\rho_p^{ss} \rangle$, the coherence terms can be subsequently deduced from $\rho_c=-\mathcal{M}_{cc}^{-1} \mathcal{M}_{cp} \rho_p$. We denote the $i^{\mathrm{th}}$ component of $|\rho_s^{ss}\rangle$ as $\rho_{ss,i}$. $|\rho_s^{ss}\rangle$ is the solution for a set of linear equations, the solutions are the rational functions. Without loss of generality, we write the $\rho_{ss,i}$ in the following form,
\beq \rho_{ss,i}=\dfrac{\mathcal{N}_i}{\mathcal{D}_i},\eeq
where $\mathcal{N}_i$ and $\mathcal{D}_i$ are polynomial functions of $\Gamma,\ \Gamma_h,\ \Delta,\ \delta_1,\ \delta_2$ and $\Omega$. Under the assumption of $\Gamma_{11}=\Gamma_{22}=\Gamma$ and the absence of thermal environment, the explicit expression of $\mathcal{N}_i$ and $\mathcal{D}_i$ are provided below.

\[
\mathcal{D}_1=\mathcal{D}_2=\mathcal{D}_3=...=\mathcal{D}_9=
\]
\begin{dmath}
-\left(\Gamma ^2+4 \Delta ^2\right) \Gamma _h^6+\left(\Gamma ^2 \delta _1^2-2 \left(5 \Gamma ^2+16 \Delta ^2\right) \delta _2 \delta _1+\Gamma ^2 \delta _2^2+\left(\Gamma ^2+4 \Delta ^2\right) \left(3 \Gamma ^2-8 \Delta ^2-4 \Omega ^2\right)\right) \Gamma _h^4+8 \Gamma  \Delta  \left(\left(2 \Gamma ^2+8 \Delta ^2+\Omega ^2\right) \delta _1+\left(2 \Gamma ^2+8 \Delta ^2-\Omega ^2\right) \delta _2\right) \Gamma _h^3-\left(3 \Gamma ^6+4 \left(3 \Delta ^2+2 \Omega ^2\right) \Gamma ^4+4 \left(4 \Delta ^4+12 \Omega ^2 \Delta ^2+\Omega ^4\right) \Gamma ^2-8 \delta _1^3 \delta _2 \Gamma ^2+2 \left(3 \Gamma ^4+4 \left(\Delta ^2+4 \Omega ^2\right) \Gamma ^2+64 \Delta ^2 \Omega ^2\right) \delta _2^2+64 \Delta ^4 \left(\Delta ^2+\Omega ^2\right)-4 \delta _1 \delta _2 \left(3 \Gamma ^4+12 \Delta ^2 \Gamma ^2+2 \delta _2^2 \Gamma ^2+32 \Delta ^4-2 \left(3 \Gamma ^2+8 \Delta ^2\right) \Omega ^2\right)+2 \delta _1^2 \left(3 \Gamma ^4+4 (\Delta -\Omega ) (\Delta +\Omega ) \Gamma ^2+16 \left(\Gamma ^2+2 \Delta ^2\right) \delta _2^2\right)\right) \Gamma _h^2-8 \Gamma  \Delta  \left(2 \Gamma ^2 \delta _1^3-2 \left(\Gamma ^2+2 \Omega ^2\right) \delta _2 \delta _1^2+\left(2 \Gamma ^4+\left(8 \Delta ^2+5 \Omega ^2\right) \Gamma ^2-2 \left(\Gamma ^2-2 \Omega ^2\right) \delta _2^2+4 (\Delta -\Omega ) \Omega ^2 (\Delta +\Omega )\right) \delta _1+\delta _2 \left(2 \Gamma ^4+2 \delta _2^2 \Gamma ^2+\left(8 \Delta ^2-5 \Omega ^2\right) \Gamma ^2-4 \Delta ^2 \Omega ^2\right)\right) \Gamma _h+\Gamma ^2 \left(4 \left(\Gamma ^2+8 \Omega ^2+4 \delta _1^2\right) \delta _2^4-8 \delta _1 \left(\Gamma ^2+4 \Delta ^2+8 \Omega ^2+4 \delta _1^2\right) \delta _2^3+\left(5 \Gamma ^4+24 \Delta ^2 \Gamma ^2+16 \Delta ^4-64 \Omega ^4+32 \left(\Gamma ^2+5 \Delta ^2\right) \Omega ^2+8 \delta _1^2 \left(3 \Gamma ^2+16 \Delta ^2+2 \delta _1^2\right)\right) \delta _2^2-2 \delta _1 \left(\Gamma ^4+24 \Delta ^2 \Gamma ^2+80 \Delta ^4-32 \Omega ^4+4 \left(\Gamma ^2-4 \Delta ^2\right) \Omega ^2+4 \left(\Gamma ^2+4 (\Delta -\Omega ) (\Delta +\Omega )\right) \delta _1^2\right) \delta _2+\left(\Gamma ^2+4 \Delta ^2+2 \Omega ^2\right) \left(16 \Omega ^4+2 \left(5 \Gamma ^2+4 \Delta ^2\right) \Omega ^2+\left(\Gamma ^2+4 \Delta ^2\right)^2\right)+\delta _1^2 \left(5 \Gamma ^4+4 \delta _1^2 \Gamma ^2+24 \left(\Delta ^2+\Omega ^2\right) \Gamma ^2+16 \left(\Delta ^4+\Omega ^4\right)\right)\right),
\end{dmath}

\begin{dmath}
\mathcal{N}_1=\mathcal{D}_1-\mathcal{N}_2-\mathcal{N}_3,
\end{dmath}

\begin{dmath*}\mathcal{N}_2=\end{dmath*}
\begin{dmath}
4 \Omega ^2 \left(\Gamma ^4 \delta _1^2-2 \Gamma ^2 \delta _1 \delta _2 \left(\Gamma ^2-4 \Omega ^2\right)+4 \Gamma ^2 \delta _2^4-8 \Gamma ^2 \delta _1 \delta _2^3+\delta _2^2 \left(\Gamma ^2 \left(5 \Gamma ^2+4 \delta _1^2+16 \Delta ^2-8 \Omega ^2\right)-4 \left(\Gamma ^2+4 \Delta ^2\right) \Gamma _h^2\right)\\+\Gamma ^2 \left(\Gamma ^2+4 \Delta ^2+2 \Omega ^2\right) \left(\Gamma ^2-\Gamma _h^2+2 \Omega ^2\right)\right),
\end{dmath}

\begin{dmath*}\mathcal{N}_3=\end{dmath*}
\begin{dmath}
4 \Omega ^2 \left(4 \Delta ^2+\Gamma _h^2\right) \left(\Gamma ^2 \left(\Gamma ^2+\left(\delta _1-\delta _2\right){}^2+4 \Delta ^2+2 \Omega ^2\right)-\left(\Gamma ^2+4 \Delta ^2\right) \Gamma _h^2\right),
\end{dmath}

\begin{dmath*}\mathcal{N}_4=\end{dmath*}
\begin{dmath}
-4 \Omega  \left(\Gamma ^2 \left(-2 \delta _2 \delta _1^2 \left(\Gamma ^2+4 \delta _2^2+2 \Delta ^2-4 \Omega ^2\right)-4 \delta _2 \Delta ^2 \left(\Gamma ^2+\delta _2^2+4 \Delta ^2-4 \Omega ^2\right)+\delta _1^3 \left(\Gamma ^2+4 \delta _2^2\right)+\delta _1 \left(\Gamma ^4+\delta _2^2 \left(5 \Gamma ^2+4 \delta _2^2+24 \Delta ^2-8 \Omega ^2\right)+4 \Gamma ^2 \left(\Delta ^2+\Omega ^2\right)+4 \Omega ^4\right)\right)+\delta _2 \left(-\left(\Gamma ^2+4 \Delta ^2\right)\right) \Gamma _h^4+2 \Gamma  \Delta  \left(\Gamma ^2+4 \Delta ^2\right) \Gamma _h^3+\Gamma _h^2 \left(\delta _2 \left(\Gamma ^2+4 \Delta ^2\right) \left(\Gamma ^2+4 \Delta ^2-4 \Omega ^2\right)+\Gamma ^2 \delta _2^3+\Gamma ^2 \delta _1^2 \delta _2-\delta _1 \left(\Gamma ^4+2 \delta _2^2 \left(3 \Gamma ^2+8 \Delta ^2\right)+4 \Gamma ^2 \Delta ^2\right)\right)-2 \Gamma  \Delta  \Gamma _h \left(\Gamma ^4+\left(\delta _1-\delta _2\right) \left(\Gamma ^2 \delta _1-\delta _2 \left(\Gamma ^2+4 \Omega ^2\right)\right)+4 \Gamma ^2 \Delta ^2-4 \Omega ^4\right)\right),
\end{dmath}

\begin{dmath*}\mathcal{N}_5=\end{dmath*}
\begin{dmath}
2 \Omega  \left(\Gamma ^3 \left(\delta _2^2 \left(5 \Gamma ^2+20 \Delta ^2-8 \Omega ^2\right)-2 \delta _1 \delta _2 \left(\Gamma ^2+4 \delta _2^2+4 \Delta ^2-4 \Omega ^2\right)+\delta _1^2 \left(\Gamma ^2+4 \delta _2^2+4 \Delta ^2\right)+\left(\Gamma ^2+4 \Delta ^2+2 \Omega ^2\right)^2+4 \delta _2^4\right)+\left(\Gamma ^3+4 \Gamma  \Delta ^2\right) \Gamma _h^4-\Gamma  \Gamma _h^2 \left(\delta _2^2 \left(5 \Gamma ^2+16 \Delta ^2\right)+\Gamma ^2 \delta _1^2-2 \Gamma ^2 \delta _1 \delta _2+2 \left(\Gamma ^2+4 \Delta ^2\right) \left(\Gamma ^2+2 \left(\Delta ^2+\Omega ^2\right)\right)\right)-8 \Gamma ^2 \Delta  \Gamma _h \left(\delta _2 \left(\Gamma ^2+4 \Delta ^2\right)+\left(\delta _1-\delta _2\right) \left(\left(\delta _1-\delta _2\right) \delta _2+\Omega ^2\right)\right)+8 \delta _2 \Delta  \left(\Gamma ^2+4 \Delta ^2\right) \Gamma _h^3\right),
\end{dmath}

\begin{dmath*}\mathcal{N}_6=\end{dmath*}
\begin{dmath}
-4 \Omega  \left(\Gamma ^2 \Delta  \left(\delta _2^2 \left(\Gamma ^2+4 \left(\Delta ^2+\Omega ^2\right)\right)+\delta _1^2 \left(\Gamma ^2+8 \delta _2^2+4 \Delta ^2-4 \Omega ^2\right)-2 \delta _1 \delta _2 \left(3 \left(\Gamma ^2+4 \Delta ^2\right)+2 \delta _2^2\right)+\left(\Gamma ^2+4 \Delta ^2\right)^2-4 \delta _1^3 \delta _2-4 \Omega ^4\right)+\Gamma  \left(\delta _1+\delta _2\right) \left(\Gamma ^2+4 \Delta ^2\right) \Gamma _h^3+\Delta  \Gamma _h^2 \left(2 \delta _1 \delta _2 \left(\Gamma ^2+8 \Delta ^2\right)+\Gamma ^2 \delta _1^2+\Gamma ^2 \delta _2^2-4 \Delta ^2 \left(\Gamma ^2+4 \Delta ^2\right)\right)-\Delta  \left(\Gamma ^2+4 \Delta ^2\right) \Gamma _h^4-\Gamma  \Gamma _h \left(\Gamma ^2 \delta _1^3-\Gamma ^2 \delta _2 \delta _1^2+\delta _1 \left(\Gamma ^4-\Gamma ^2 \delta _2^2+4 \Gamma ^2 \left(\Delta ^2+\Omega ^2\right)+8 \Delta ^2 \Omega ^2\right)+\delta _2 \left(\Gamma ^4+\Gamma ^2 \delta _2^2-4 \Omega ^2 \left(\Gamma ^2+2 \Delta ^2\right)+4 \Gamma ^2 \Delta ^2\right)\right)\right),
\end{dmath}

\begin{dmath*}\mathcal{N}_7=\end{dmath*}
\begin{dmath}
2 \Omega  \left(-4 \Gamma ^3 \Delta  \left(\delta _1 \left(\Gamma ^2-\delta _2^2+4 \left(\Delta ^2+\Omega ^2\right)\right)+\delta _2 \left(\Gamma ^2+\delta _2^2+4 \Delta ^2-4 \Omega ^2\right)+\delta _1^3-\delta _2 \delta _1^2\right)+4 \Gamma  \Delta  \Gamma _h^2 \left(\delta _1 \left(\Gamma ^2+4 \Delta ^2+2 \Omega ^2\right)+\delta _2 \left(\Gamma ^2+4 \Delta ^2-2 \Omega ^2\right)\right)-\left(\Gamma ^2+4 \Delta ^2\right) \Gamma _h^5+\Gamma ^2 \Gamma _h \left(-\Gamma ^4-\delta _2^2 \left(\Gamma ^2-4 \Delta ^2+4 \Omega ^2\right)-\delta _1^2 \left(\Gamma ^2+8 \delta _2^2-4 \left(\Delta ^2+\Omega ^2\right)\right)+2 \delta _1 \delta _2 \left(3 \Gamma ^2+2 \delta _2^2+4 \Delta ^2\right)+4 \delta _1^3 \delta _2+16 \Delta ^4+4 \Omega ^4\right)+\Gamma _h^3 \left(-2 \delta _2 \delta _1 \left(3 \Gamma ^2+8 \Delta ^2\right)+\Gamma ^2 \delta _1^2+\Gamma ^2 \delta _2^2+2 \left(\Gamma ^4+2 \Gamma ^2 \Delta ^2-8 \Delta ^4\right)\right)\right),
\end{dmath}

\begin{dmath*}\mathcal{N}_8=\end{dmath*}
\begin{dmath}
-4 \Omega ^2 \left(4 \Gamma ^2 \Delta  \left(\delta _2 \left(\Gamma ^2+4 \Delta ^2\right)+\left(\delta _1-\delta _2\right) \left(\left(\delta _1-\delta _2\right) \delta _2+\Omega ^2\right)\right)-\left(\Gamma ^3+4 \Gamma  \Delta ^2\right) \Gamma _h^3-4 \delta _2 \Delta  \left(\Gamma ^2+4 \Delta ^2\right) \Gamma _h^2+\Gamma _h \left(\Gamma ^3 \left(\delta _1-\delta _2\right){}^2+\Gamma  \left(\Gamma ^2+4 \Delta ^2\right) \left(\Gamma ^2+2 \Omega ^2\right)\right)\right),
\end{dmath}

\begin{dmath*}\mathcal{N}_9=\end{dmath*}
\begin{dmath}
8 \Omega ^2 \left(\Gamma ^3 \Delta  \left(\Gamma ^2+\left(\delta _1-\delta _2\right){}^2+4 \Delta ^2+2 \Omega ^2\right)-\Gamma ^2 \Gamma _h \left(\delta _2 \left(\Gamma ^2+4 \Delta ^2\right)+\left(\delta _1-\delta _2\right) \left(\left(\delta _1-\delta _2\right) \delta _2+\Omega ^2\right)\right)+\delta _2 \left(\Gamma ^2+4 \Delta ^2\right) \Gamma _h^3-\Gamma  \Delta  \Gamma _h^2 \left(\Gamma ^2+4 \Delta ^2+2 \Omega ^2\right)\right).\end{dmath}

\subsection*{Rabi splitting with detuning}

\begin{figure}[htbp!]
\centering
{\includegraphics[width=.45\columnwidth]{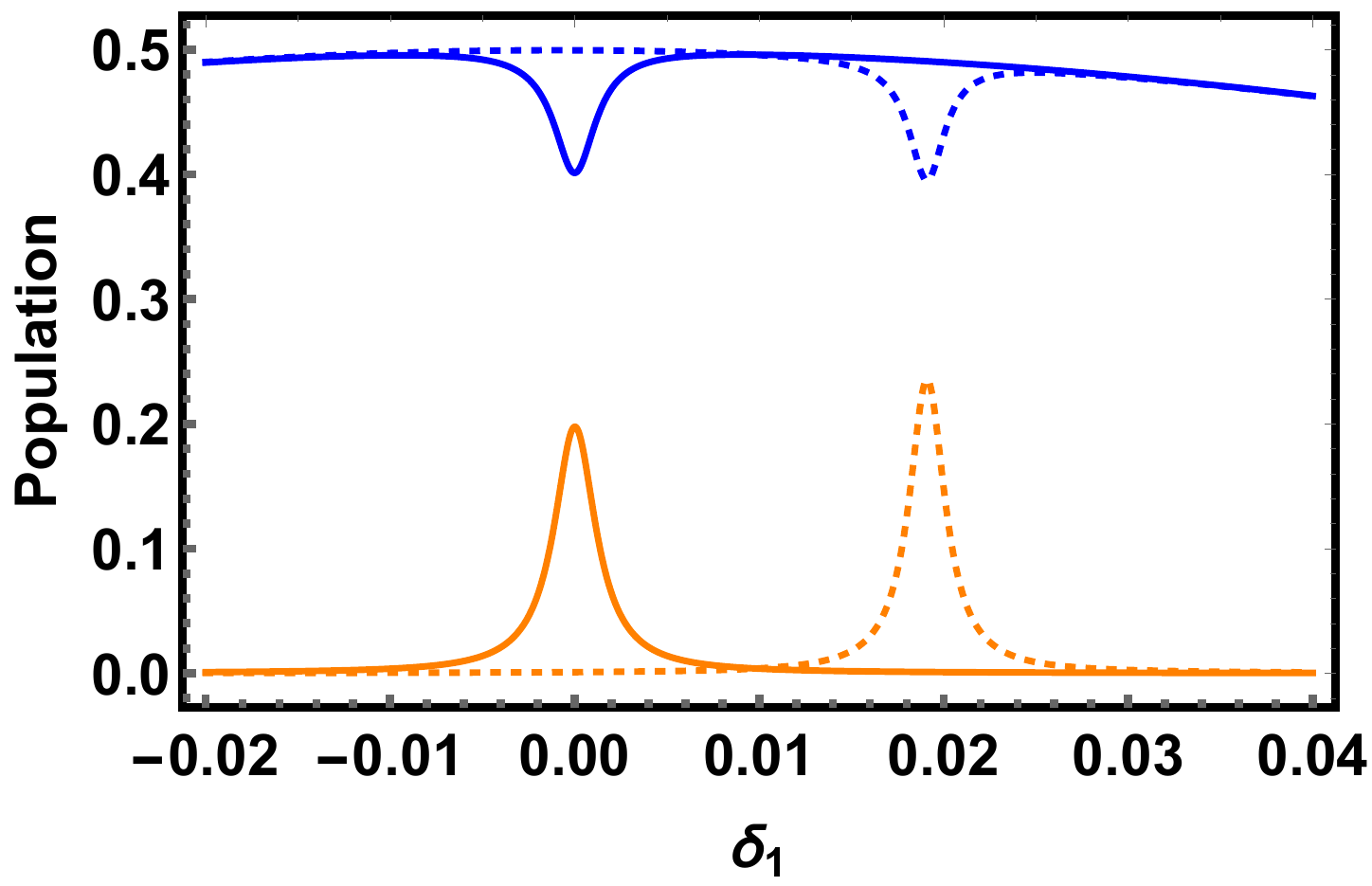}}
\caption{Population on state ``a'' (navy, top) and ``b'' (orange, bottom) vs detuning $\delta_1$. Solid lines: $\Omega=R^*=0.1$. Dotted lines: $\Omega=0.1$, $R^*=0.11$. The peak value of excitation of around 0.2 can be inferred from the constant term in Eqn.~\ref{kappar}. Here $\Delta=5\times 10^{-4}$, $\Gamma=10^{-4}$.}
    \label{detuning}
\end{figure}

We consider the more general case when the frequency of the driven radiation is slightly tuned away from the donor frequency. Then, the spitted energy levels due to the dynamic stark effect will be shifted to $\frac{\nu+\omega_1}{2}\pm \delta \omega$ where $\delta \omega = \sqrt{\Omega^2+\frac{\delta_1^2}{4}}$ [Ref.~\cite{kiffner}]. When one of the split energy levels is in resonant with that of the acceptor, i.e. $\omega_2=\frac{\nu+\omega_1}{2}\pm \delta \omega$, the population on acceptor will increase dramatically. This happens when $\delta_1=\frac{\delta_2^2-\Omega^2}{\delta_2}$. For example, when $\Omega=0.1$, and $\delta_2=\delta_1-0.1$, the resonance happens at zero detuning $\delta_1=0$ [Fig.~\ref{detuning} solid lines]. The resonance of the dressed states leads to a faster transportation of excitations from the donor to the acceptor, therefore we observe an unusual dip on the navy line in Fig.~\ref{detuning}. A more general scenario when the Rabi frequency $\Omega$ is not exactly equal to the energy splitting between the two states is plotted in dotted lines. From the above argument, the peak happens at around 0.019 which is consistent with the plot.

\end{widetext}

\bigskip

\end{document}